%% file: main.tex
\documentclass[preliminary,copyright]{eptcs}
\providecommand{\event}{MARS 2015}
\usepackage{breakurl}
\usepackage{rotating}
\usepackage{graphicx}
\usepackage{subfig}
\usepackage[usenames,dvipsnames]{xcolor}
\definecolor{grey}{cmyk}{0,0,0,0.8}
\hypersetup{bookmarks=true,bookmarksopen=false,colorlinks=true,linkcolor=grey,citecolor=grey,urlcolor=grey}

\makeatletter
\newcommand\Paragraph{\@startsection{paragraph}{4}{\z@}%
                                    {1.5ex \@plus1ex \@minus.2ex}%
                                    {-1em}%
                                    {\normalfont\normalsize\bfseries}}
\makeatother

\input{macr_pic}

\usepackage{listings}
\usepackage{cadp-lnt}
\usepackage{cadp-lotos}
\usepackage{cadp-svl}
\title{Formal Specification and Verification of Fully Asynchronous Implementations of the Data Encryption Standard}
\author{Wendelin Serwe
\institute{Inria}
\institute{Univ. Grenoble Alpes, LIG, F-38000 Grenoble, France}
\institute{CNRS, LIG, F-38000 Grenoble, France}
\email{wendelin.serwe@inria.fr}
}
\def\titlerunning{Asynchronous Data Encryption Standard}
\def\authorrunning{W. Serwe}
\begin{document}
\maketitle

\begin{abstract}
  This paper presents two formal models of the Data Encryption Standard (DES), a first using the international standard LOTOS, and a second using the more recent process calculus LNT.
  Both models encode the DES in the style of asynchronous circuits, i.e., the data-flow blocks of the DES algorithm are represented by processes communicating via rendezvous.
  To ensure correctness of the models, several techniques have been applied, including model checking, equivalence checking, and comparing the results produced by a prototype automatically generated from the formal model with those of existing implementations of the DES.
  The complete code of the models is provided as appendices and also available on the website of the CADP verification toolbox.
\end{abstract}

\section{Introduction}

The Data Encryption Standard (DES) is a symmetric-key encryption algorithm, that has been for almost 30 years a Federal Information Processing Standard~\cite{FIPS-46-3}.
At present, the main weakness of the algorithm is its use of keys of 56 bits, which are too short to withstand brute force attacks.
To address this issue, the Triple Data Encryption Algorithm (TDEA or Triple DES) variant~\cite{NIST-SP-800-67} applies the DES algorithm three times (encryption, decryption, encryption), using three different keys, and is still an approved symmetric cryptographic algorithm for 8-byte block ciphers, e.g., for secure payment systems~\cite[Annex B1.1]{EMV-4.3}.

An interesting aspect of the DES is that it is specified by a data-flow diagram, i.e., as a set of blocks communicating by message passing.
Such an architecture is naturally asynchronous because there is no need for a global clock synchronizing these various blocks.
Due to different interleavings of the block executions, the exists a risk that the DES execution produces a nondeterministic result, although this is not expected.
This problem naturally lends itself to analysis with process calculi.

A prior case-study~\cite{Borrione-Boubekeur-Mounier-et-al-06} analyzed an asynchronous circuit%
\footnote{An asynchronous circuit is a circuit without clocks, where the different components of the circuit synchronize via handshake protocols.
  Implementing a cryptographic algorithm as an asynchronous circuit makes it more robust against side channel attacks based on the analysis of the power consumption or radio-emission, which have significantly fewer peaks that could be exploited to get information about the key or data being encrypted~\cite{Moore-Anderson-Cunnigham-et-al-03}.}
implementing the DES.
This circuit was specified in the CHP (Communicating Hardware Processes)~\cite{Martin-86} process calculus, which was translated into the IF~\cite{Bozga-Fernandez-Ghirvu-et-al-99-a} formalism, and subsequently verified using the CADP toolbox~\cite{Garavel-Lang-Mateescu-Serwe-13}\footnote{\url{http://cadp.inria.fr}}, i.e., the corresponding state space was generated as an LTS (Labeled Transition System) and several properties were verified by model and equivalence checking.
To reduce the state space, this prior case-study replaced some of the parallelism by sequential compositions, removing some asynchronism.

The present paper goes further and analyzes the fully asynchronous DES, describing formal models of the DES in two different process calculi, namely the international standard LOTOS~\cite{ISO-8807} and the more recent LNT language~\cite{Champelovier-Clerc-Garavel-et-al-10-v6.3}, both supported by CADP.
The LOTOS model of the DES was directly derived from the DES standard~\cite{FIPS-46-3} to experiment with the application of LOTOS for the analysis of asynchronous circuits.
In August 2015, the LOTOS model was rewritten into an equivalent LNT model to fulfill the expectations of the MARS workshop.
Both models have been analyzed using different techniques, namely data abstraction, model checking, equivalence checking, and the automatic generation of a prototype software implementation.
All these verification steps have been automated by an SVL (Script Verification Language)~\cite{Garavel-Lang-01} script.

The rest of this paper is organized as follows.
Section~\ref{sec:architecture} briefly presents the data-flow architecture of the DES.
Section~\ref{sec:modeling} discusses modeling challenges and choices.
Section~\ref{sec:abstract} (respectively, \ref{sec:concrete}) presents the steps undertaken to assess the correctness of the model with (respectively without) data abstraction.
Section~\ref{sec:conclusion} gives concluding remarks.
Appendix~\ref{sec:lnt} is the complete source code of the LNT model.
Appendix~\ref{sec:lotos} is the complete source code of the LOTOS model.
Appendix~\ref{sec:execcaesar_c} contains the C code required to generate a prototype with the EXEC/C\AE{}SAR framework.
Appendix~\ref{sec:svl} provides the SVL script to execute the verification scenarios described in this paper.

\section{Architecture of the Asynchronous Data Encryption Standard}
\label{sec:architecture}

The DES is an iterative algorithm, which takes as input a 64-bit word\footnote{Longer data must be split into 64-bit words, possibly adding zeros at the end so that the total number of bits is a multiple of 64.  Each 64-bit word is then encrypted or decrypted separately.} of data, a 64-bit key (from which only 56 bits are used), and a bit indicating whether the data is to be encrypted or decrypted, and produces as output a 64-bit word of encrypted or decrypted data.
The DES first applies an initial permutation on the input data, splits the 64-bit data word into two 32-bit data words, named $L_0$ and $R_0$, and then iteratively computes $L_{n+1} = R_n$ and $R_{n+1} = L_n \oplus f(R_n, K_{n+1})$, where $\oplus$ stands for the bit-wise sum, and $f$ is the so-called \emph{cipher function}.
The final result is obtained as the inverse initial permutation applied to the concatenation of $R_{16}$ and $L_{16}$.

In each iteration, the cipher function $f$ first applies a function named E to expand the 32-bit word $R_n$ to a 48-bit word, and then computes the bit-wise sum with the 48-bit subkey $K_n$.
The result is split into eight 6-bit words, each of which is transformed by a so-called S-box into a 4-bit word.
The output of the cipher function is the concatenation of these eight 4-bit words, permuted by a function named P.

To compute the sixteen subkeys $K_n$ (with $n \in \{1, ..., 16\}$), a function named PC1 (permuted choice) selects 56 bits from the 64-bit key, and then splits the result into two 28-bit words, named $C_0$ and $D_0$.
For iteration $n$, the subkey $K_n$ is obtained by application of a function named PC2 to select 48 bits from $C_nD_n$, which are defined by successive shift operations according to a schedule defined in the standard.
Decryption follows the same scheme, only applying the subkeys in the opposite order.


\section{Formal Models of the Asynchronous Data Encryption Standard}
\label{sec:modeling}

\begin{sidewaysfigure}
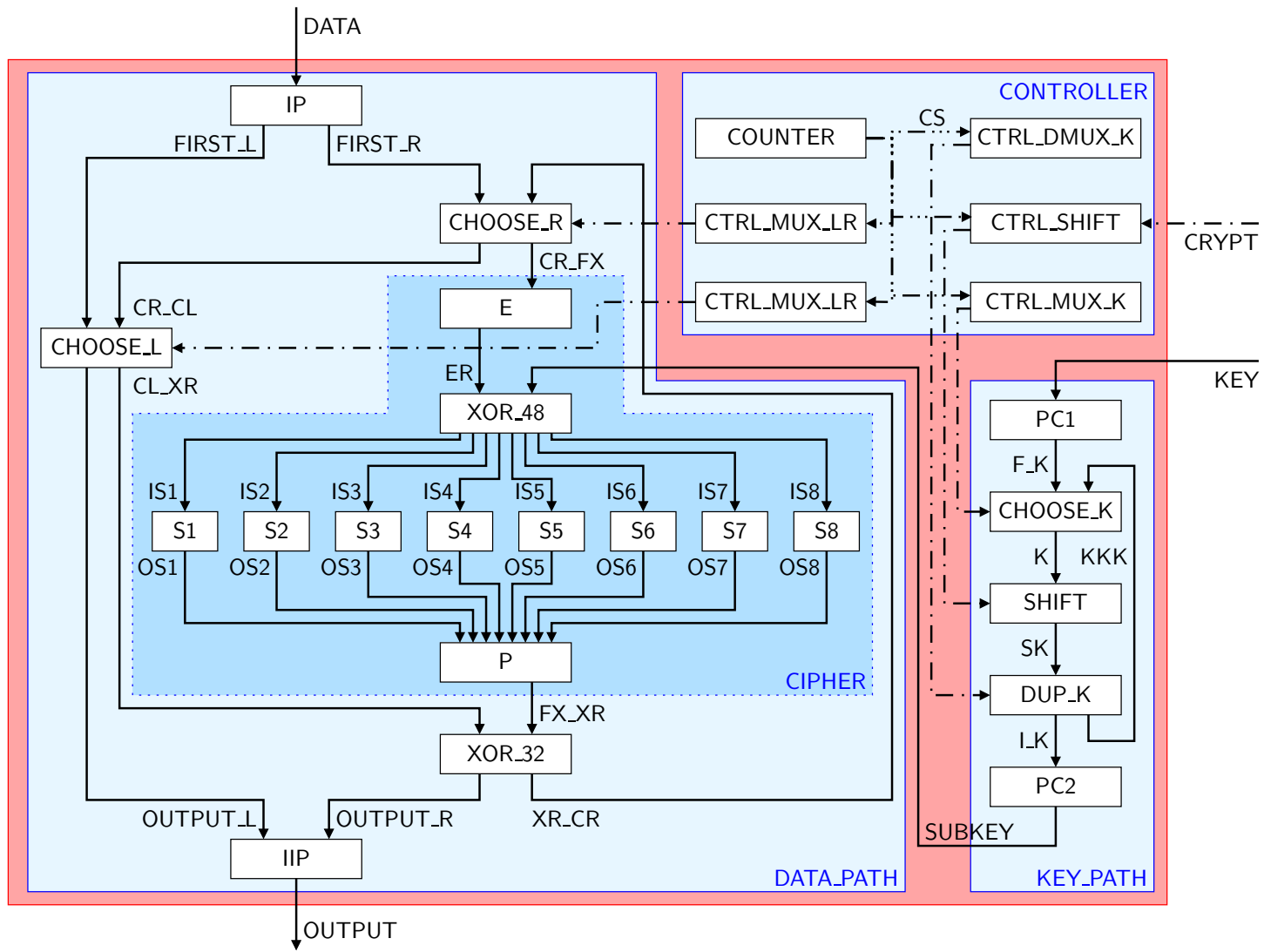

  \Figure{architecture}
  \vspace*{-2ex}
  \caption{Architecture of the asynchronous Data Encryption Standard.
    Boxes represent processes, and arrows indicate synchronizations (dashed arrows correspond to the control signals).
    Arrows are labeled by the name of the corresponding gate, where \lstinline[language=LNT]+F_K+ stands for \lstinline[language=LNT]+FIRST_K+, and \lstinline[language=LNT]+I_K+ stands for \lstinline[language=LNT]+INTERMEDIATE_K+.}
  \label{fig:architecture}
\end{sidewaysfigure}

Figure~\ref{fig:architecture} shows the overall architecture of the LOTOS and LNT models, which closely follow the architecture of the standard.
Each block (permutation, S-box, bit-wise sum, shift, etc.) is represented by a process, communicating by rendezvous with its neighbors.
Thus, there is no need for a global clock: each block waits for its operands, performs its operation, and transmits the result to the subsequent block.

In addition to the processes with a direct correspondence to blocks of the DES standard, the models include ten processes without a direct correspondence, but required to share blocks (or processes) among all iterations of the algorithm: the six processes of the \lstinline[language=LNT]+CONTROLLER+, and the four processes \lstinline[language=LNT]+CHOOSE_L+, \lstinline[language=LNT]+CHOOSE_R+, \lstinline[language=LNT]+CHOOSE_K+, and \lstinline[language=LNT]+DUP_K+.
The latter four are arbiters and/or multiplexers, which select among several inputs and possibly duplicate their output.

The process \lstinline[language=LNT]+COUNTER+ generates the control signal ``\mbox{\lstinline[language=LNT]+CS !+$i$}'', where $i$ starts from $0$ and is increased up to $16$, thus indicating seventeen steps (one more than the number of iterations in the algorithm).
The signal \lstinline[language=LNT]+CS+ is sent to the five other processes of the \lstinline[language=LNT]+CONTROLLER+, which in turn indicate to the shift register the number and direction of the shift(s) and to the four arbiters which input to read from and/or which output to write to.
Seventeen steps are required because the processes \lstinline[language=LNT]+CHOOSE_L+ and \lstinline[language=LNT]+CHOOSE_R+ take as input either the initial data or the result of the previous iteration, and output either to the next iteration or the final output.
Hence it is necessary to execute them before and after the sixteen iterations, explaining the additional step.
Precisely, the rendezvous ``\mbox{\lstinline[language=LNT]+CS !16+}'' triggers a rendezvous only in process \lstinline[language=LNT]+CTRL_MUX_LR+, whereas it is simply consumed by \lstinline[language=LNT]+CTRL_MUX_K+, \lstinline[language=LNT]+CTRL_DMUX_K+, and \lstinline[language=LNT]+CTRL_SHIFT+.

Notice that the formal models contain no non-deterministic choice (\lstinline[language=LNT]+select+ operator of LNT or ``\lstinline[language=LOTOS]+[]+'' operator of LOTOS), because every block of the DES is deterministic.

The complete LOTOS and LNT models are given as appendices and also available on the CADP website\footnote{\url{http://cadp.inria.fr/demos/demo_38}}.
\newcounter{demo38}\setcounter{demo38}{\value{footnote}}
Table~\ref{tab:lines} gives the number of lines for the two models, plus those of the LOTOS model generated by the LNT2LOTOS translator from the LNT model.
The LNT model is significantly shorter and syntactically closer to the DES standard.
In particular, the definition of the S-boxes using tables (as in the standard) is more convenient, due to the automatically defined functions to manipulate arrays.
Compared to the hand-written models, the generated LOTOS model is much larger due to many automatically generated functions and auxiliary processes that are ``inlined'' in the hand-written models.

Rewriting the LOTOS model into LNT uncovered a few errors in the LOTOS definitions of the S-boxes and a small bug\footnote{See entry \#2076 in the list of changes to CADP (\url{http://cadp.inria.fr/changes.html}).} in the LNT2LOTOS translator.
This rewrite also showed that the controller was too restrictive.
Precisely, the five control processes (\lstinline[language=LNT]+CTRL_MUX_K+, \lstinline[language=LNT]+CTRL_SHIFT+, \lstinline[language=LNT]+CTRL_DMUX_K+, and the two instances of \lstinline[language=LNT]+CTRL_MUX_LR+) were synchronized at the end of an encoding or decoding, although \lstinline[language=LNT]+CTRL_SHIFT+ could accept a new input on gate \lstinline[language=LNT]+CRYPT+ as soon as the shift-command for the generation of the last subkey has been sent to the shift register.%
\footnote{Removing this unnecessary synchronization (also present in the CHP and IF models) slightly increases the LTS size: the initial LOTOS model yielded an LTS with 588,785,433 states and 5,512,418,012 transitions.}

\section{Analysis of the Abstract Model}
\label{sec:abstract}

Analyzing the DES with enumerative techniques is challenging, because it is unfeasible to enumerate all 64-bit vectors.
A first approach is to abstract from the actual \emph{data} values and to focus the analysis on the control part, e.g., check whether the sixteen iterations are correctly synchronized.
Redefining the \lstinline[language=LNT]+BIT+ data type to contain a single value (instead of two values), automatically transforms all bit-vector types into singleton types and all functions operating on bit vectors to the identity function.\footnote{The IF model~\cite{Borrione-Boubekeur-Mounier-et-al-06} used a similar abstraction.}
Due to this drastic abstraction, enumerating all possible inputs is trivial, enabling LTS generation without resorting to environments and compositional techniques~\cite{Garavel-Lang-Mateescu-15}.

\begin{table}
  \newcommand{\M}[1]{\makebox[2em][r]{#1}}
  \newcommand{\I}[1]{\M{\it#1}}
  \small\centering
  \begin{tabular}{lcccclrr}
    \cline{1-4} \cline{6-8}
    & LOTOS & LNT & gen.~LOTOS & \hspace{2em} &
    & \multicolumn{1}{c}{LOTOS} & \multicolumn{1}{c}{~~LNT} \\ 
    \cline{1-4} \cline{6-8}
    types \& functions & \M{1172} &  \M{575} & \M{2514} && states      &   591,914,192 &   167,300,852 \\
    channels           &    \M{0} &   \M{50} &   \M{58} && transitions & 5,542,917,498 & 1,500,073,686 \\
    processes          &  \M{671} &  \M{668} &  \M{772} && time (min)  &           228 &            66 \\
    \cline{1-4}
    {\it total}        & \I{1843} & \I{1293} & \I{3344} && RAM (GB)    &         19.13 &          4.93 \\
    \cline{1-4} \cline{6-8}
    \multicolumn{4}{c}{\parbox{.4\textwidth}{\caption{Line number count of the models}\label{tab:lines}}} &&
    \multicolumn{3}{c}{\parbox{.4\textwidth}{\caption{Direct LTS Generation}\label{tab:lts}\centering\vspace*{-1.5ex}(on an Xeon(R) E5-2630 at 2.4 GHz)\footnotemark{}}}
  \end{tabular}
\end{table}
\footnotetext{
  The significantly smaller size (5.3 million states and 30 million transitions) of the LTS corresponding to the IF model~\cite{Borrione-Boubekeur-Mounier-et-al-06} is explained by the fact that in the IF model the eight S-boxes execute sequentially rather than in parallel~\cite[page~305, footnote~4]{Salaun-Serwe-05}.
  Applying a similar restriction to the LNT model, the size of the corresponding LTS drops to 1,375,048 states and 7,804,352 transitions (respectively to 8,183,770 states and 45,025,227 transitions for a restricted LOTOS model).
  A further difference is that the controller in the IF model is a single automaton and enforces a stronger ordering between the inputs and outputs than the more asynchronous architecture shown in Figure~\ref{fig:architecture}.
}

\Paragraph{Direct LTS Generation and Equivalence between the LNT and LOTOS Models.}

Table~\ref{tab:lts} gives statistics about the LTSs generated directly from the abstract models.
The two models (LOTOS and LNT) yield LTSs equivalent for branching (but not strong) bisimulation (checked by \lstinline[language=SVL]+PROPERTY_7+ of the SVL script given in Appendix~\ref{sec:svl}).
In both cases, the LTS minimized for branching bisimulation has 28 states and 78 transitions.

The significant difference in LTS size between the LNT model and the LOTOS model is explained by the semantic difference in the symmetric sequential composition~\cite{Garavel-15-b}.
The LOTOS operator ``\lstinline[language=LOTOS]+>>+'' generates an internal ``\lstinline[language=LOTOS]+i+'' transition, whereas the LNT operator ``\lstinline[language=LNT]+;+'' does not.
Such a symmetric sequential composition (rather than a simple action prefix) is required whenever a process can read several inputs in parallel before producing its output.
This is in particular the case for process \lstinline[language=LNT]+P+, which reads the output of the eight S-boxes.
When these internal ``\lstinline[language=LOTOS]+i+'' transitions are removed by adding the pragma ``\lstinline[language=LOTOS]+(*! atomic *)+'' to all occurrences of ``\lstinline[language=LOTOS]+>>+'', the LOTOS model yields an LTS of exactly the same size as the LNT model.

\begin{figure}[tp]
  \centering
  \includegraphics[bb=240 40 550 490, scale=.55]{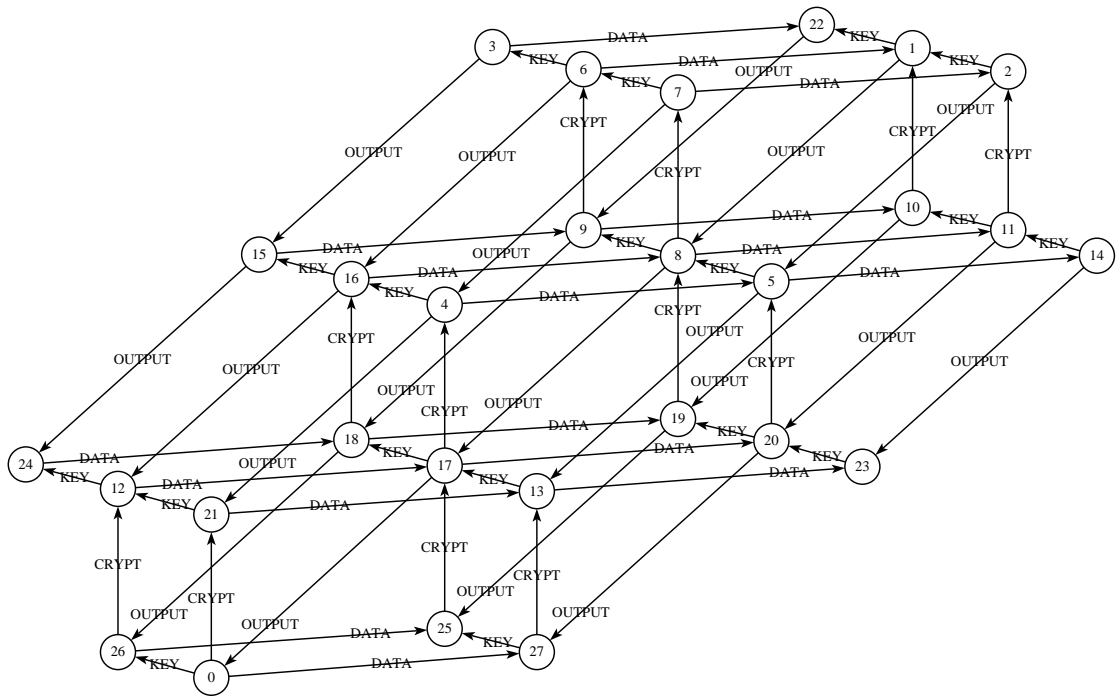}
  \caption{LTS of the abstract DES.
    All offers have been stripped from the transition labels.
    Thus, a transition labeled with \texttt{CRYPT} represents two transitions, labeled ``\lstinline[language=LOTOS]+CRYPT !false+'' and ``\lstinline[language=LOTOS]+CRYPT !true+''.
    For all other transitions, the offer corresponds to the single value in the domain of abstract 64-bit vectors.}
  \label{fig:abstract_LTS}
\end{figure}

The LTS obtained after removing all offers (using appropriate renaming operations, i.e., those applied to file \lstinline[language=SVL]+des_sample.bcg+ in \lstinline[language=SVL]+PROPERTY_6+ of the SVL script given in Appendix~\ref{sec:svl}) and minimization using branching bisimulation is shown in Figure~\ref{fig:abstract_LTS}.

\Paragraph{Compositional LTS Generation.}

Compositional LTS generation, i.e., the bottom-up LTS construction alternating generation and minimization steps, is much more efficient.
The initial steps of the SVL script given in Appendix~\ref{sec:svl} require at most 50 MB of RAM and generate the minimized LTS (28 states and 78 transitions) in less than 90 seconds.
The success of compositional techniques on the LOTOS model triggered the development of the CHP2LOTOS translator~\cite{Garavel-Salaun-Serwe-09} to provide the full power of CADP to the designers of asynchronous circuits.

\Paragraph{Model Checking.}

Several properties of the control part have been analyzed formally, using model checking and equivalence checking on the LTS generated compositionally from the abstract models.

A first property, called \lstinline[language=SVL]+PROPERTY_1+ in the SVL script, expresses the absence of deadlocks.

A second property, called \lstinline[language=SVL]+PROPERTY_2+ expresses the fact that a triplet of inputs on gates \lstinline[language=LNT]+CRYPT+, \lstinline[language=LNT]+DATA+, and \lstinline[language=LNT]+KEY+ is eventually followed by an rendezvous on gate \lstinline[language=LNT]+OUTPUT+.

A third property, called \lstinline[language=SVL]+PROPERTY_3+, describes the asynchronism of the DES models, i.e., that the DES may accept the inputs for $N$ future rounds before it produces the result of the current round.
This property is expressed by two temporal logic formul\ae, a first one expressing that the DES never accepts more than $N$ inputs in advance, and a second one expressing that there exists an execution, where the DES indeed accepts $N$ inputs in advance.
This third property is parametric, because $N$ varies for the three inputs \lstinline[language=LNT]+DATA+, \lstinline[language=LNT]+CRYPT+, and \lstinline[language=LNT]+KEY+.


A last property, called \lstinline[language=SVL]+PROPERTY_4+, expresses the correct synchronization between the data path and the key path, namely that each encoding or decoding executes the sixteen iterations.
This property can be verified in two ways: by model checking a temporal logic formula and by checking the equivalence of the generated labeled transition system with a simple automaton containing a loop of sixteen \lstinline[language=LNT]+SUBKEY+ transitions (because there must be a subkey per iteration) interleaved with a \lstinline[language=LNT]+CRYPT+ transition (because each encoding or decoding requires one such transition).
The LNT (respectively, LOTOS) model of this automaton is given in Appendix~\ref{sec:lnt:property_4} (respectively, \ref{sec:lotos:property_4}).
Notice that this verification requires to keep the \lstinline[language=LNT]+SUBKEY+ gate visible.

\section{Analysis of the Concrete Model}
\label{sec:concrete}

The concrete models (i.e., without data abstraction) can be used to check the correctness of the results computed by the DES.
Two different approaches were used: model checking and the generation of a prototype implementation.

\Paragraph{Rapid Prototyping.}
\label{sec:execcaesar}

Using the EXEC/C\AE{}SAR framework~\cite{Garavel-Viho-Zendri-01} for rapid prototyping, it is possible to generate a C implementation of the LOTOS or LNT model: one only has to provide C functions implementing the interaction with the environment on the visible gates \lstinline[language=LNT]+CRYPT+, \lstinline[language=LNT]+DATA+, \lstinline[language=LNT]+KEY+, and \lstinline[language=LNT]+OUTPUT+.
This C code is given in Appendix~\ref{sec:execcaesar_c}.

We first checked that the prototype is reversible, i.e., that deciphering a cipher with the same key results in the original data.
These tests helped to spot and correct some errors in the subkey generation.

We also compared the results of the prototypes (LOTOS and LNT), and to those of some other publicly available implementations\footnote{See for instance \url{https://www.schneier.com/books/applied_cryptography/source.html}}.
These comparisons helped to spot and correct a handful of differences caused by typographic errors, i.e., bad copying of the DES standard into LOTOS and LNT.
The compilation and execution of the prototype implementation corresponds to \lstinline[language=SVL]+PROPERTY_5+ in the SVL script given in Appendix~\ref{sec:svl}.
Although using a handful of tests is far from exhaustive, the structure of the DES with its iterations, permutations and bit operations and the fact that cipher heavily shuffles the input data lead to a coverage sufficient for the purpose of analyzing the \emph{control} part.

\Paragraph{Model Checking.}

The generation of an LTS from the concrete model requires an environment to restrict the domain of possible input values.
The model \lstinline[language=LNT]+DES_SAMPLE+ is such a variant of the concrete one, including a sequential environment providing a key and a data for a single encryption and checking the correctness of the output.
The direct generation of the corresponding LTS (10,156,715 states and 75,933,635 transitions for the LNT model, and 32,219,740 states and 259,010,596 transitions for the LOTOS model) requires less than 6~GB of RAM.

The correctness of the computed result can verified by model checking a property expressing that the action on gate \lstinline[language=LNT]+OUTPUT+ is eventually reached.
A second verification is that after removing all offers, the LTS is included in the LTS of Figure~\ref{fig:abstract_LTS}.
These verifications are carried out by the \lstinline[language=SVL]+PROPERTY_6+ in the SVL script given in Appendix~\ref{sec:svl}.

\section{Conclusion}
\label{sec:conclusion}

This paper presents two formal models of the Data Encryption Standard, which might be an interesting benchmark example, because it is both complex and tractable (with current hardware and/or compositional techniques).
For instance, it is sufficiently large to make interesting screenshots of the monitor of the distributed state space generation tool DISTRIBUTOR~\cite{Garavel-Mateescu-Bergamini-et-al-06} (see also the DISTRIBUTOR manual page\footnote{\url{http://cadp.inria.fr/man/distributor.html\#sect7}}).
These models also illustrate an interesting feature of LOTOS and LNT, namely the possibility to easily change the implementation of a data type to transform a prototype implementation into an abstract model adapted to formal verification.

{\small
\paragraph{Acknowledgements.}

Some of the experiments presented in this paper were carried out using the Grid'5000 experimental testbed\footnote{See \url{http://www.grid5000.fr}} built by Inria with support from CNRS, RENATER, several Universities, and other funding bodies.
I am grateful to Hubert Garavel for his suggestion to improve the model(s) and their presentation.
I would like to thank Edith Beign\'e, Fran\c{c}ois Bertrand, Pascal Vivet (CEA Leti), Dominique Borrione, Menouer Boubekeur, Marc Renaudin (TIMA), and Gwen Sala\"un (Inria) for discussions about asynchronous logic, the CHP language, and the implementation of the DES in asynchronous logic.
}

\input{main.bbl}
\appendix

\input{lnt.tex}

\input{lotos.tex}

\input{c.tex}

\input{svl.tex}

\end{document}

%% file: macr_pic.tex
\input{psfig}

\newlength{\magic}

\makeatletter
\def\Picture[#1]{\@ifnextchar[{\@lPicture[#1]}{\@Picture[#1]}}
\def\@Picture[#1]#2{\par\begin{center}\@lPicture[#1][14]{#2}\end{center}\par}
\def\@lPicture[#1][#2]#3{\mbox{\vspace*{1ex}\hspace*{0.8cm}\psfig{rheight=#1cm,rwidth=#2cm,bbllx=3.1714cm,bblly=0cm,bburx=0cm,bbury=\magic,figure=#3.ps}\hspace*{-0.8cm}\vspace*{1ex}}}
\makeatother

\makeatletter
\def\Figure{\@ifnextchar[{\@lFigure}{\@Figure}}
\def\@Figure#1{\begin{center} \input{#1} \end{center}}
\def\@lFigure[#1]#2{\hspace*{#1}{\parbox{\textwidth}{\@Figure{#2}}}}
\makeatother

%% file: lnt.tex
\section{LNT Model of an Asynchronous DES}
\label{sec:lnt}

\lstset{language=LNT}

This appendix gives the complete LNT model of the DES, as required by the SVL verification scenario given in Appendix~\ref{sec:svl}.
This model requires CADP version 2015-h ``Stony Brook'' (August 2015) or later.
The syntax and semantics of LNT is defined in the reference manual~\cite{Champelovier-Clerc-Garavel-et-al-10-v6.3}.

\subsection{Module BIT\_CONCRETE}

\begin{lstlisting}[firstline=5]
module BIT_CONCRETE is

type BIT is
   -- bit data type with two different values
   0 !implementedby "BIT_ZERO",
   1 !implementedby "BIT_ONE"
   with "=="
end type

end module

\end{lstlisting}

\subsection{Module BIT\_ABSTRACT}
\label{sec:lnt:abstract_bit}

This module defines the abstraction replacing the concrete implementation of type \lstinline+BIT+ (see module \lstinline+BIT_CONCRETE.lnt+ above), transforming the two-valued type into a singleton, which transitively transforms all bit vectors into singletons as well, thus completely abstracting data.

\begin{lstlisting}[firstline=5]
module BIT_ABSTRACT is

type BIT is
   -- bit data type abstracted to a single value
   0
   with "=="
end type

function 1 : BIT is
   return 0
end function

end module

\end{lstlisting}

\subsection{Module TYPES}
\label{sec:lnt:types}

\begin{lstlisting}[firstline=5]
module TYPES (BIT) with "get" is

---------------------------------------------------------------------------
-- exclusive-or operation on (abstract or concrete) bits

function _xor_ (X, Y : BIT) : BIT is
   if X == Y then
      return 0
   else
      return 1
   end if
end function

---------------------------------------------------------------------------
-- data type for counting the iterations of the algorithm

type ITERATION is
   range 0 .. 16 of NAT
   with "==", "!=", "first", "last"
end type

---------------------------------------------------------------------------
-- data type for the controlling the behavior of the shift register

type SHIFT is
   NO,  -- no shift
   LS1, -- 1 left shift
   LS2, -- 2 left shifts
   RS1, -- 1 right shift
   RS2  -- 2 right shifts
   with "=="
end type

---------------------------------------------------------------------------
-- data type for the control signals used for the multiplexers

type PHASE is
   F, -- first iteration
   N, -- intermediate iteration
   L  -- last iteration
   with "=="
end type

---------------------------------------------------------------------------
-- 2-bit vectors

type BIT2 is
   MK_2 (B1, B2: BIT)
end type

---------------------------------------------------------------------------
-- conversion of a 2-bit vector into a natural number
-- note: the most significant bit is B1

function BIT2_TO_NAT (X: BIT2) : NAT is
   var N: NAT in
      N := 0;
      if X.B1 == 1 then N := N + 2 end if;
      if X.B2 == 1 then N := N + 1 end if;
      return N
   end var
end function

---------------------------------------------------------------------------
-- 4-bit vectors

type BIT4 is !implementedby "ADT_BIT4"
   MK_4 (B1, B2, B3, B4: BIT) !implementedby "MK_4"
end type

---------------------------------------------------------------------------
-- conversion of a 4-bit vector into a natural number
-- note: the most significant bit is B1

function BIT4_TO_NAT (X: BIT4) : NAT is
   var N: NAT in
      N := 0;
      if X.B1 == 1 then N := N + 8 end if;
      if X.B2 == 1 then N := N + 4 end if;
      if X.B3 == 1 then N := N + 2 end if;
      if X.B4 == 1 then N := N + 1 end if;
      return N
   end var
end function

---------------------------------------------------------------------------
-- conversion of a natural number into a 4-bit vector
-- note: the most significant bit is B1

function NAT_TO_BIT4 (N: NAT) : BIT4 is
   case N in
      0      -> return MK_4 (0, 0, 0, 0)
   |  1      -> return MK_4 (0, 0, 0, 1)
   |  2      -> return MK_4 (0, 0, 1, 0)
   |  3      -> return MK_4 (0, 0, 1, 1)
   |  4      -> return MK_4 (0, 1, 0, 0)
   |  5      -> return MK_4 (0, 1, 0, 1)
   |  6      -> return MK_4 (0, 1, 1, 0)
   |  7      -> return MK_4 (0, 1, 1, 1)
   |  8      -> return MK_4 (1, 0, 0, 0)
   |  9      -> return MK_4 (1, 0, 0, 1)
   | 10      -> return MK_4 (1, 0, 1, 0)
   | 11      -> return MK_4 (1, 0, 1, 1)
   | 12      -> return MK_4 (1, 1, 0, 0)
   | 13      -> return MK_4 (1, 1, 0, 1)
   | 14      -> return MK_4 (1, 1, 1, 0)
   | any NAT -> return MK_4 (1, 1, 1, 1)
   end case
end function

---------------------------------------------------------------------------
-- 6-bit vectors

type BIT6 is
   MK_6 (B1, B2, B3, B4, B5, B6: BIT)
end type

---------------------------------------------------------------------------
-- projection of 6-bit vectors to 2-bit vectors

function 1AND6 (X: BIT6) : BIT2 is
   return MK_2 (X.B1, X.B6)
end function

---------------------------------------------------------------------------
-- projection of 6-bit vectors to 4-bit vectors

function 2TO5 (X: BIT6) : BIT4 is
   return MK_4 (X.B2, X.B3, X.B4, X.B5)
end function

---------------------------------------------------------------------------
-- 32-bit vectors

type BIT32 is
   MK_32 (B1,  B2,  B3,  B4,  B5,  B6,  B7,  B8,
          B9,  B10, B11, B12, B13, B14, B15, B16,
          B17, B18, B19, B20, B21, B22, B23, B24,
          B25, B26, B27, B28, B29, B30, B31, B32: BIT)
end type

---------------------------------------------------------------------------
-- bitwise XOR for 32-bit vectors

function XOR (A, B: BIT32) : BIT32 is
   return MK_32 (A.B1  xor B.B1,  A.B2  xor B.B2,
                 A.B3  xor B.B3,  A.B4  xor B.B4,
                 A.B5  xor B.B5,  A.B6  xor B.B6,
                 A.B7  xor B.B7,  A.B8  xor B.B8,
                 A.B9  xor B.B9,  A.B10 xor B.B10,
                 A.B11 xor B.B11, A.B12 xor B.B12,
                 A.B13 xor B.B13, A.B14 xor B.B14,
                 A.B15 xor B.B15, A.B16 xor B.B16,
                 A.B17 xor B.B17, A.B18 xor B.B18,
                 A.B19 xor B.B19, A.B20 xor B.B20, 
                 A.B21 xor B.B21, A.B22 xor B.B22,
                 A.B23 xor B.B23, A.B24 xor B.B24,
                 A.B25 xor B.B25, A.B26 xor B.B26,
                 A.B27 xor B.B27, A.B28 xor B.B28,
                 A.B29 xor B.B29, A.B30 xor B.B30,
                 A.B31 xor B.B31, A.B32 xor B.B32)
end function
   
---------------------------------------------------------------------------
-- concatenation of eight 4-bit vectors to form a 32-bit vector

function MK_32 (V1, V2, V3, V4, V5, V6, V7, V8: BIT4) : BIT32 is
   return MK_32 (V1.B1, V1.B2, V1.B3, V1.B4,  V2.B1, V2.B2, V2.B3, V2.B4,
                 V3.B1, V3.B2, V3.B3, V3.B4,  V4.B1, V4.B2, V4.B3, V4.B4,
                 V5.B1, V5.B2, V5.B3, V5.B4,  V6.B1, V6.B2, V6.B3, V6.B4,
                 V7.B1, V7.B2, V7.B3, V7.B4,  V8.B1, V8.B2, V8.B3, V8.B4)
end function

---------------------------------------------------------------------------
-- 48-bit vectors

type BIT48 is
   MK_48 (B1,  B2,  B3,  B4,  B5,  B6,  B7,  B8,
          B9,  B10, B11, B12, B13, B14, B15, B16,
          B17, B18, B19, B20, B21, B22, B23, B24,
          B25, B26, B27, B28, B29, B30, B31, B32,
          B33, B34, B35, B36, B37, B38, B39, B40,
          B41, B42, B43, B44, B45, B46, B47, B48: BIT)
end type

---------------------------------------------------------------------------
-- bitwise XOR for 48-bit vectors

function XOR (A, B: BIT48) : BIT48 is
   return MK_48 (A.B1  xor B.B1,  A.B2  xor B.B2,
                 A.B3  xor B.B3,  A.B4  xor B.B4, 
                 A.B5  xor B.B5,  A.B6  xor B.B6,
                 A.B7  xor B.B7,  A.B8  xor B.B8,
                 A.B9  xor B.B9,  A.B10 xor B.B10,
                 A.B11 xor B.B11, A.B12 xor B.B12,
                 A.B13 xor B.B13, A.B14 xor B.B14,
                 A.B15 xor B.B15, A.B16 xor B.B16,
                 A.B17 xor B.B17, A.B18 xor B.B18,
                 A.B19 xor B.B19, A.B20 xor B.B20,
                 A.B21 xor B.B21, A.B22 xor B.B22,
                 A.B23 xor B.B23, A.B24 xor B.B24,
                 A.B25 xor B.B25, A.B26 xor B.B26,
                 A.B27 xor B.B27, A.B28 xor B.B28,
                 A.B29 xor B.B29, A.B30 xor B.B30,
                 A.B31 xor B.B31, A.B32 xor B.B32,
                 A.B33 xor B.B33, A.B34 xor B.B34,
                 A.B35 xor B.B35, A.B36 xor B.B36,
                 A.B37 xor B.B37, A.B38 xor B.B38,
                 A.B39 xor B.B39, A.B40 xor B.B40, 
                 A.B41 xor B.B41, A.B42 xor B.B42,
                 A.B43 xor B.B43, A.B44 xor B.B44,
                 A.B45 xor B.B45, A.B46 xor B.B46,
                 A.B47 xor B.B47, A.B48 xor B.B48)
end function

---------------------------------------------------------------------------
-- projections of 48-bit vectors to 6-bit vectors

function 1TO6 (X: BIT48) : BIT6 is
   return MK_6 (X.B1, X.B2, X.B3, X.B4, X.B5, X.B6)
end function

function 7TO12 (X: BIT48) : BIT6 is
   return MK_6 (X.B7, X.B8, X.B9, X.B10, X.B11, X.B12)
end function

function 13TO18 (X: BIT48) : BIT6 is
   return MK_6 (X.B13, X.B14, X.B15, X.B16, X.B17, X.B18)
end function

function 19TO24 (X: BIT48) : BIT6 is
   return MK_6 (X.B19, X.B20, X.B21, X.B22, X.B23, X.B24)
end function

function 25TO30 (X: BIT48) : BIT6 is
   return MK_6 (X.B25, X.B26, X.B27, X.B28, X.B29, X.B30)
end function

function 31TO36 (X: BIT48) : BIT6 is
   return MK_6 (X.B31, X.B32, X.B33, X.B34, X.B35, X.B36)
end function

function 37TO42 (X: BIT48) : BIT6 is
   return MK_6 (X.B37, X.B38, X.B39, X.B40, X.B41, X.B42)
end function

function 43TO48 (X: BIT48) : BIT6 is
   return MK_6 (X.B43, X.B44, X.B45, X.B46, X.B47, X.B48)
end function

---------------------------------------------------------------------------
-- 56-bit vectors

type BIT56 is
   MK_56 (B1,  B2,  B3,  B4,  B5,  B6,  B7,  B8,
          B9,  B10, B11, B12, B13, B14, B15, B16,
          B17, B18, B19, B20, B21, B22, B23, B24,
          B25, B26, B27, B28, B29, B30, B31, B32,
          B33, B34, B35, B36, B37, B38, B39, B40,
          B41, B42, B43, B44, B45, B46, B47, B48,
          B49, B50, B51, B52, B53, B54, B55, B56: BIT)
end type

---------------------------------------------------------------------------
-- left shift of a 56-bit vector
-- note: more precisely, parallel left shift of two 28-bit vectors

function LSHIFT (X: BIT56) : BIT56 is
   return MK_56 (X.B2,  X.B3,  X.B4,  X.B5,  X.B6,  X.B7,  X.B8,  X.B9,
                 X.B10, X.B11, X.B12, X.B13, X.B14, X.B15, X.B16, X.B17,  
                 X.B18, X.B19, X.B20, X.B21, X.B22, X.B23, X.B24, X.B25, 
                 X.B26, X.B27, X.B28, X.B1,  X.B30, X.B31, X.B32, X.B33, 
                 X.B34, X.B35, X.B36, X.B37, X.B38, X.B39, X.B40, X.B41, 
                 X.B42, X.B43, X.B44, X.B45, X.B46, X.B47, X.B48, X.B49, 
                 X.B50, X.B51, X.B52, X.B53, X.B54, X.B55, X.B56, X.B29)
end function

---------------------------------------------------------------------------
-- right shift of a 56-bit vector
-- note: more precisely, parallel right shift of two 28-bit vectors

function RSHIFT (X: BIT56) : BIT56 is
   return MK_56 (X.B28, X.B1,  X.B2,  X.B3,  X.B4,  X.B5,  X.B6,  X.B7,
                 X.B8,  X.B9,  X.B10, X.B11, X.B12, X.B13, X.B14, X.B15,
                 X.B16, X.B17, X.B18, X.B19, X.B20, X.B21, X.B22, X.B23,
                 X.B24, X.B25, X.B26, X.B27, X.B56, X.B29, X.B30, X.B31,
                 X.B32, X.B33, X.B34, X.B35, X.B36, X.B37, X.B38, X.B39,
                 X.B40, X.B41, X.B42, X.B43, X.B44, X.B45, X.B46, X.B47,
                 X.B48, X.B49, X.B50, X.B51, X.B52, X.B53, X.B54, X.B55)
end function

---------------------------------------------------------------------------
-- 64-bit vectors

type BIT64 is !implementedby "ADT_BIT64" !printedby "ADT_PRINT_BIT64"
   MK_64 (B1,  B2,  B3,  B4,  B5,  B6,  B7,  B8,
          B9,  B10, B11, B12, B13, B14, B15, B16,
          B17, B18, B19, B20, B21, B22, B23, B24,
          B25, B26, B27, B28, B29, B30, B31, B32,
          B33, B34, B35, B36, B37, B38, B39, B40,
          B41, B42, B43, B44, B45, B46, B47, B48,
          B49, B50, B51, B52, B53, B54, B55, B56,
          B57, B58, B59, B60, B61, B62, B63, B64: BIT)
   !implementedby "MK_64"
   with "==", "!="
end type

---------------------------------------------------------------------------
-- projections of 64-bit vectors to 32-bit vectors

function 1TO32 (X: BIT64) : BIT32 is
   return MK_32 (X.B1,  X.B2,  X.B3,  X.B4,  X.B5,  X.B6,  X.B7,  X.B8,
                 X.B9,  X.B10, X.B11, X.B12, X.B13, X.B14, X.B15, X.B16,
                 X.B17, X.B18, X.B19, X.B20, X.B21, X.B22, X.B23, X.B24,
                 X.B25, X.B26, X.B27, X.B28, X.B29, X.B30, X.B31, X.B32)
end function

function 33TO64 (X: BIT64) : BIT32 is
   return MK_32 (X.B33, X.B34, X.B35, X.B36, X.B37, X.B38, X.B39, X.B40,
                 X.B41, X.B42, X.B43, X.B44, X.B45, X.B46, X.B47, X.B48,
                 X.B49, X.B50, X.B51, X.B52, X.B53, X.B54, X.B55, X.B56,
                 X.B57, X.B58, X.B59, X.B60, X.B61, X.B62, X.B63, X.B64)
end function

---------------------------------------------------------------------------
-- concatenation of sixteen 4-bit vectors to form a 64-bit vector

function MK_64 (V1, V2,  V3,  V4,  V5,  V6,  V7,  V8,
                V9, V10, V11, V12, V13, V14, V15, V16: BIT4) : BIT64 is
   !implementedby "CONCAT_BIT4"
      return MK_64 (V1.B1,  V1.B2,  V1.B3,  V1.B4,
                    V2.B1,  V2.B2,  V2.B3,  V2.B4,
                    V3.B1,  V3.B2,  V3.B3,  V3.B4,
                    V4.B1,  V4.B2,  V4.B3,  V4.B4,
                    V5.B1,  V5.B2,  V5.B3,  V5.B4,
                    V6.B1,  V6.B2,  V6.B3,  V6.B4,
                    V7.B1,  V7.B2,  V7.B3,  V7.B4,
                    V8.B1,  V8.B2,  V8.B3,  V8.B4,
                    V9.B1,  V9.B2,  V9.B3,  V9.B4, 
                    V10.B1, V10.B2, V10.B3, V10.B4,
                    V11.B1, V11.B2, V11.B3, V11.B4,
                    V12.B1, V12.B2, V12.B3, V12.B4,
                    V13.B1, V13.B2, V13.B3, V13.B4,
                    V14.B1, V14.B2, V14.B3, V14.B4,
                    V15.B1, V15.B2, V15.B3, V15.B4,
                    V16.B1, V16.B2, V16.B3, V16.B4)
end function

---------------------------------------------------------------------------
-- concatenation of two 32-bit vectors to form a 64-bit vector

function MK_64 (V1, V2: BIT32) : BIT64 is
   return MK_64 (V1.B1,  V1.B2,  V1.B3,  V1.B4,
                 V1.B5,  V1.B6,  V1.B7,  V1.B8,
                 V1.B9,  V1.B10, V1.B11, V1.B12,
                 V1.B13, V1.B14, V1.B15, V1.B16,
                 V1.B17, V1.B18, V1.B19, V1.B20,
                 V1.B21, V1.B22, V1.B23, V1.B24,
                 V1.B25, V1.B26, V1.B27, V1.B28,
                 V1.B29, V1.B30, V1.B31, V1.B32,
                 V2.B1,  V2.B2,  V2.B3,  V2.B4,
                 V2.B5,  V2.B6,  V2.B7,  V2.B8,
                 V2.B9,  V2.B10, V2.B11, V2.B12,
                 V2.B13, V2.B14, V2.B15, V2.B16,
                 V2.B17, V2.B18, V2.B19, V2.B20,
                 V2.B21, V2.B22, V2.B23, V2.B24,
                 V2.B25, V2.B26, V2.B27, V2.B28,
                 V2.B29, V2.B30, V2.B31, V2.B32)
end function

end module

\end{lstlisting}

\subsection{Module PERMUTATION\_FUNCTIONS}

\begin{lstlisting}[firstline=5]
module PERMUTATION_FUNCTIONS (TYPES) is

---------------------------------------------------------------------------
-- E: expansion of a 32-bit vector to a 48-bit vector

function E (X: BIT32) : BIT48 is
   return MK_48 (X.B32, X.B1,  X.B2,  X.B3,  X.B4,  X.B5,
                 X.B4,  X.B5,  X.B6,  X.B7,  X.B8,  X.B9,
                 X.B8,  X.B9,  X.B10, X.B11, X.B12, X.B13,
                 X.B12, X.B13, X.B14, X.B15, X.B16, X.B17,
                 X.B16, X.B17, X.B18, X.B19, X.B20, X.B21,
                 X.B20, X.B21, X.B22, X.B23, X.B24, X.B25,
                 X.B24, X.B25, X.B26, X.B27, X.B28, X.B29,
                 X.B28, X.B29, X.B30, X.B31, X.B32, X.B1)
end function

---------------------------------------------------------------------------
-- IP: initial permutation

function IP (X: BIT64) : BIT64 is
   return MK_64 (X.B58, X.B50, X.B42, X.B34, X.B26, X.B18, X.B10, X.B2,
                 X.B60, X.B52, X.B44, X.B36, X.B28, X.B20, X.B12, X.B4,
                 X.B62, X.B54, X.B46, X.B38, X.B30, X.B22, X.B14, X.B6,
                 X.B64, X.B56, X.B48, X.B40, X.B32, X.B24, X.B16, X.B8,
                 X.B57, X.B49, X.B41, X.B33, X.B25, X.B17, X.B9,  X.B1,
                 X.B59, X.B51, X.B43, X.B35, X.B27, X.B19, X.B11, X.B3,
                 X.B61, X.B53, X.B45, X.B37, X.B29, X.B21, X.B13, X.B5,
                 X.B63, X.B55, X.B47, X.B39, X.B31, X.B23, X.B15, X.B7)
end function

---------------------------------------------------------------------------
-- IIP: inverse initial permutation

function IIP (X: BIT64) : BIT64 is
   return MK_64 (X.B40, X.B8,  X.B48, X.B16, X.B56, X.B24, X.B64, X.B32,
                 X.B39, X.B7,  X.B47, X.B15, X.B55, X.B23, X.B63, X.B31,
                 X.B38, X.B6,  X.B46, X.B14, X.B54, X.B22, X.B62, X.B30,
                 X.B37, X.B5,  X.B45, X.B13, X.B53, X.B21, X.B61, X.B29,
                 X.B36, X.B4,  X.B44, X.B12, X.B52, X.B20, X.B60, X.B28,
                 X.B35, X.B3,  X.B43, X.B11, X.B51, X.B19, X.B59, X.B27,
                 X.B34, X.B2,  X.B42, X.B10, X.B50, X.B18, X.B58, X.B26,
                 X.B33, X.B1,  X.B41, X.B9,  X.B49, X.B17, X.B57, X.B25)
end function

---------------------------------------------------------------------------
-- P: permutation

function P (X: BIT32) : BIT32 is
   return MK_32 (X.B16, X.B7,  X.B20, X.B21,
                 X.B29, X.B12, X.B28, X.B17,
                 X.B1,  X.B15, X.B23, X.B26,
                 X.B5,  X.B18, X.B31, X.B10,
                 X.B2,  X.B8,  X.B24, X.B14,
                 X.B32, X.B27, X.B3,  X.B9,
                 X.B19, X.B13, X.B30, X.B6,
                 X.B22, X.B11, X.B4,  X.B25)
end function

---------------------------------------------------------------------------
-- PC1: permuted choice 1

function PC1 (X: BIT64) : BIT56 is
   return MK_56 (X.B57, X.B49, X.B41, X.B33, X.B25, X.B17, X.B9,
                 X.B1,  X.B58, X.B50, X.B42, X.B34, X.B26, X.B18,
                 X.B10, X.B2,  X.B59, X.B51, X.B43, X.B35, X.B27,
                 X.B19, X.B11, X.B3,  X.B60, X.B52, X.B44, X.B36,
   
                 X.B63, X.B55, X.B47, X.B39, X.B31, X.B23, X.B15,
                 X.B7,  X.B62, X.B54, X.B46, X.B38, X.B30, X.B22,
                 X.B14, X.B6,  X.B61, X.B53, X.B45, X.B37, X.B29,
                 X.B21, X.B13, X.B5,  X.B28, X.B20, X.B12, X.B4)
end function

---------------------------------------------------------------------------
-- PC2: permuted choice 2

function PC2 (X: BIT56) : BIT48 is
   return MK_48 (X.B14, X.B17, X.B11, X.B24, X.B1,  X.B5,
                 X.B3,  X.B28, X.B15, X.B6,  X.B21, X.B10,
                 X.B23, X.B19, X.B12, X.B4,  X.B26, X.B8,
                 X.B16, X.B7,  X.B27, X.B20, X.B13, X.B2,
                 X.B41, X.B52, X.B31, X.B37, X.B47, X.B55,
                 X.B30, X.B40, X.B51, X.B45, X.B33, X.B48,
                 X.B44, X.B49, X.B39, X.B56, X.B34, X.B53,
                 X.B46, X.B42, X.B50, X.B36, X.B29, X.B32)
end function

end module

\end{lstlisting}

\subsection{Module S\_BOX\_FUNCTIONS}
\label{sec:lnt:s_boxes}

The LNT model directly encodes the S-box tables as two-dimensional arrays, requiring the definition of the additional data types \lstinline+ROW+ and \lstinline+S_BOX_ARRAY+, together with accessor functions \lstinline+GET_ROW()+ and \lstinline+GET_COLUMN()+ and projection functions \lstinline+1AND6()+ and \lstinline+2TO5()+ (the latter are part of module \lstinline+TYPES+ (see Appendix~\ref{sec:lnt:types}).

\begin{lstlisting}[firstline=5]
module S_BOX_FUNCTIONS (TYPES) is

---------------------------------------------------------------------------
-- data types defining a two-dimensional array to encode the S-boxes

type ROW is
   array [0..15] of NAT
end type

type S_BOX_ARRAY is
   array [0..3] of ROW
end type

---------------------------------------------------------------------------

function GET_ROW (X: BIT6) : NAT is
   return BIT2_TO_NAT (1AND6 (X))
end function

---------------------------------------------------------------------------

function GET_COLUMN (X: BIT6) : NAT is
   return BIT4_TO_NAT (2TO5 (X))
end function

---------------------------------------------------------------------------

function S1 : S_BOX_ARRAY is
   return S_BOX_ARRAY
     (ROW (14,  4, 13,  1,  2, 15, 11,  8,  3, 10,  6, 12,  5,  9,  0,  7),
      ROW ( 0, 15,  7,  4, 14,  2, 13,  1, 10,  6, 12, 11,  9,  5,  3,  8),
      ROW ( 4,  1, 14,  8, 13,  6,  2, 11, 15, 12,  9,  7,  3, 10,  5,  0),
      ROW (15, 12,  8,  2,  4,  9,  1,  7,  5, 11,  3, 14, 10,  0,  6, 13))
end function

---------------------------------------------------------------------------

function S2 : S_BOX_ARRAY is
   return S_BOX_ARRAY 
     (ROW (15,  1,  8, 14,  6, 11,  3,  4,  9,  7,  2, 13, 12,  0,  5, 10),
      ROW ( 3, 13,  4,  7, 15,  2,  8, 14, 12,  0,  1, 10,  6,  9, 11,  5),
      ROW ( 0, 14,  7, 11, 10,  4, 13,  1,  5,  8, 12,  6,  9,  3,  2, 15),
      ROW (13,  8, 10,  1,  3, 15,  4,  2, 11,  6,  7, 12,  0,  5, 14,  9))
end function

---------------------------------------------------------------------------

function S3 : S_BOX_ARRAY is
   return S_BOX_ARRAY 
     (ROW (10,  0,  9, 14,  6,  3, 15,  5,  1, 13, 12,  7, 11,  4,  2,  8),
      ROW (13,  7,  0,  9,  3,  4,  6, 10,  2,  8,  5, 14, 12, 11, 15,  1),
      ROW (13,  6,  4,  9,  8, 15,  3,  0, 11,  1,  2, 12,  5, 10, 14,  7),
      ROW ( 1, 10, 13,  0,  6,  9,  8,  7,  4, 15, 14,  3, 11,  5,  2, 12))
end function

---------------------------------------------------------------------------

function S4 : S_BOX_ARRAY is
   return S_BOX_ARRAY 
     (ROW ( 7, 13, 14,  3,  0,  6,  9, 10,  1,  2,  8,  5, 11, 12,  4, 15),
      ROW (13,  8, 11,  5,  6, 15,  0,  3,  4,  7,  2, 12,  1, 10, 14,  9),
      ROW (10,  6,  9,  0, 12, 11,  7, 13, 15,  1,  3, 14,  5,  2,  8,  4),
      ROW ( 3, 15,  0,  6, 10,  1, 13,  8,  9,  4,  5, 11, 12,  7,  2, 14))
end function

---------------------------------------------------------------------------

function S5 : S_BOX_ARRAY is
   return S_BOX_ARRAY 
     (ROW ( 2, 12,  4,  1,  7, 10, 11,  6,  8,  5,  3, 15, 13,  0, 14,  9),
      ROW (14, 11,  2, 12,  4,  7, 13,  1,  5,  0, 15, 10,  3,  9,  8,  6),
      ROW ( 4,  2,  1, 11, 10, 13,  7,  8, 15,  9, 12,  5,  6,  3,  0, 14),
      ROW (11,  8, 12,  7,  1, 14,  2, 13,  6, 15,  0,  9, 10,  4,  5,  3))
end function

---------------------------------------------------------------------------

function S6 : S_BOX_ARRAY is
   return S_BOX_ARRAY
     (ROW (12,  1, 10, 15,  9,  2,  6,  8,  0, 13,  3,  4, 14,  7,  5, 11),
      ROW (10, 15,  4,  2,  7, 12,  9,  5,  6,  1, 13, 14,  0, 11,  3,  8),
      ROW ( 9, 14, 15,  5,  2,  8, 12,  3,  7,  0,  4, 10,  1, 13, 11,  6),
      ROW ( 4,  3,  2, 12,  9,  5, 15, 10, 11, 14,  1,  7,  6,  0,  8, 13))
end function

---------------------------------------------------------------------------

function S7 : S_BOX_ARRAY is
   return S_BOX_ARRAY
     (ROW ( 4, 11,  2, 14, 15,  0,  8, 13,  3, 12,  9,  7,  5, 10,  6,  1),
      ROW (13,  0, 11,  7,  4,  9,  1, 10, 14,  3,  5, 12,  2, 15,  8,  6),
      ROW ( 1,  4, 11, 13, 12,  3,  7, 14, 10, 15,  6,  8,  0,  5,  9,  2),
      ROW ( 6, 11, 13,  8,  1,  4, 10,  7,  9,  5,  0, 15, 14,  2,  3, 12))
end function

---------------------------------------------------------------------------

function S8 : S_BOX_ARRAY is
   return S_BOX_ARRAY
     (ROW (13,  2,  8,  4,  6, 15, 11,  1, 10,  9,  3, 14,  5,  0, 12,  7),
      ROW ( 1, 15, 13,  8, 10,  3,  7,  4, 12,  5,  6, 11,  0, 14,  9,  2),
      ROW ( 7, 11,  4,  1,  9, 12, 14,  2,  0,  6, 10, 13, 15,  3,  5,  8),
      ROW ( 2,  1, 14,  7,  4, 10,  8, 13, 15, 12,  9,  0,  3,  5,  6, 11))
end function

end module

\end{lstlisting}

\subsection{Module CHANNELS}

\begin{lstlisting}[firstline=5]
module CHANNELS (TYPES) is

---------------------------------------------------------------------------
-- channel types for bit vectors

channel C4 is
   (BIT4)
end channel

channel C6 is
   (BIT6)
end channel

channel C32 is
   (BIT32)
end channel

channel C48 is
   (BIT48)
end channel

channel C56 is
   (BIT56)
end channel

channel C64 is
   (BIT64)
end channel

---------------------------------------------------------------------------
-- channel types for scalar data types

channel CB is
   (BOOL)
end channel

channel CIT is
   (ITERATION)
end channel

channel CP is
   (PHASE)
end channel

channel CS is
   (SHIFT)
end channel

end module

\end{lstlisting}

\subsection{Module CONTROLLER}

\begin{lstlisting}[firstline=5]
module CONTROLLER (CHANNELS) is

---------------------------------------------------------------------------
-- CONTROLLER executes five concurrent processes controlling the various
-- multiplexers and the shift register, which are synchronized via a sixth
-- process counting the iterations
 
process CONTROLLER [CRYPT: CB,
                    CTRL_CL, CTRL_CR: CP,
                    CTRL_SHIFT: CS, CTRL_DK, CTRL_CK: CP] is
   hide CS: CIT in
      par CS in
         COUNTER [CS]
      ||
         CTRL_MUX_LR [CS, CTRL_CL]
      ||
         CTRL_MUX_LR [CS, CTRL_CR]
      ||
         CTRL_SHIFT [CRYPT, CS, CTRL_SHIFT]
      ||
         CTRL_DMUX_K [CS, CTRL_DK]
      ||
         CTRL_MUX_K [CS, CTRL_CK]
      end par
   end hide
end process

---------------------------------------------------------------------------
-- COUNTER counts the iterations (modulo 17)

process COUNTER [CS: CIT] is
   var IT: ITERATION in
      IT := 0;
      loop
	 CS (IT);
	 IT := ITERATION (((NAT (IT) + 1) mod 17))
      end loop
   end var
end process

---------------------------------------------------------------------------
-- CTRL_MUX_LR generates 1 F, then 15 Ns, and finally 1 L to control the 
-- multiplexers in the data path

process CTRL_MUX_LR [CS: CIT, CTRL: CP] is
   var IT: ITERATION in
      loop
         CS (?IT);
	 if IT == (0 of ITERATION) then
	    CTRL (F)
	 elsif IT == last then
	    CTRL (L)
	 else
	    CTRL (N)
	 end if
      end loop
   end var
end process

---------------------------------------------------------------------------
-- CTRL_SHIFT controls the shift register

process CTRL_SHIFT [CRYPT: CB, CS: CIT, CTRL: CS] is
   var CRYPT: BOOL, IT: ITERATION in
      loop
	 CRYPT (?CRYPT);
         loop LL in
            CS (?IT);
            if IT == last then
               break LL
	    elsif CRYPT then
	       if (IT == (0 of ITERATION)) or (IT == (1 of ITERATION)) or
                  (IT == (8 of ITERATION)) or (IT == (15 of ITERATION))
               then
		  CTRL (LS1)
	       else
		  CTRL (LS2)
	       end if
            else
	       if IT == (0 of ITERATION) then
		  CTRL (NO)
	       elsif (IT == (1 of ITERATION)) or
                     (IT == (8 of ITERATION)) or
                     (IT == (15 of ITERATION))
               then
		  CTRL (RS1)
	       else
		  CTRL (RS2)
	       end if
            end if
         end loop
      end loop
   end var
end process

---------------------------------------------------------------------------
-- CTRL_DMUX_K generates 15 Ns and 1 L to control the doubling multiplexer

process CTRL_DMUX_K [CS: CIT, CTRL: CP] is
   var IT: ITERATION in
      loop
         CS (?IT);
	 if IT == (15 of ITERATION) then
	    CTRL (L) 
	 elsif IT != last then
	    CTRL (N)
         end if
      end loop
   end var
end process

---------------------------------------------------------------------------
-- CTRL_MUX_K generates 1 F and 15 Ns to control the key path multiplexer

process CTRL_MUX_K [CS: CIT, CTRL: CP] is
   var IT: ITERATION in
      loop
         CS (?IT);
         if IT == (0 of ITERATION) then
            CTRL (F)
         elsif IT != last then
            CTRL (N)
         end if
      end loop
   end var
end process

end module

\end{lstlisting}

\subsection{Module KEY\_PATH}

\begin{lstlisting}[firstline=5]
module KEY_PATH (CHANNELS) is

---------------------------------------------------------------------------
-- KEY_PATH generates the 16 subkeys of the key schedule

process KEY_PATH [KEY: C64, SUBKEY: C48,
                  CTRL_SHIFT: CS, CTRL_DK, CTRL_CK: CP] is
   hide FIRST_K, INTERMEDIATE_K: C56 in
      par
         FIRST_K ->
            PC1 [KEY, FIRST_K]
      ||
         FIRST_K, INTERMEDIATE_K ->
            hide K, KKK, SK: C56 in
               par
                  K, SK ->
                     SHIFT_REGISTER [CTRL_SHIFT, K, SK]
               ||
                  SK, KKK ->
                     DUPLICATE_K [CTRL_DK, SK, INTERMEDIATE_K, KKK]
               ||
                  KKK, K ->
                     CHOOSE_K [CTRL_CK, FIRST_K, KKK, K]
               end par
            end hide   
      ||
         INTERMEDIATE_K ->
            PC2 [INTERMEDIATE_K, SUBKEY]
      end par
   end hide
end process

---------------------------------------------------------------------------
-- PC1 applies PC1 to select a 56-bit vector from the initial 64-bit key

process PC1 [KEY: C64, FIRST_K: C56] is
   var K64: BIT64 in
      loop
	 KEY (?K64);
	 FIRST_K (PC1 (K64))
      end loop
   end var
end process

---------------------------------------------------------------------------
-- SHIFT_REGISTER performs, depending on the iteration, one or two shifts
-- to the left or right of a 56-bit word

process SHIFT_REGISTER [CTRL: CS, INPUT, OUTPUT: C56] is
   var CTRL: SHIFT, I56: BIT56 in
      loop
         par
            CTRL (?CTRL)
         ||
            INPUT (?I56)
         end par;
         case CTRL in
           NO ->
               OUTPUT (I56)
         | LS1 ->
               OUTPUT (LSHIFT (I56))
         | LS2 ->
               OUTPUT (LSHIFT (LSHIFT (I56)))
         | RS1 ->
               OUTPUT (RSHIFT (I56))
         | RS2 ->
               OUTPUT (RSHIFT (RSHIFT (I56)))
         end case
      end loop
   end var
end process

---------------------------------------------------------------------------
-- DUPLICATE_K reads a 56-bit vector from INPUT and outputs it to OUTPUT1
-- and OUTPUT2, but for the last iteration, where it outputs only to
-- OUTPUT1. Because DUPLICATE_K always reads from INPUT, the order of the
-- rendezvous on CTRL and INPUT is arbitrary.

process DUPLICATE_K [CTRL: CP, INPUT, OUTPUT1, OUTPUT2: C56] is
   var CTRL: PHASE, I56:BIT56 in
      loop
         par
            CTRL (?CTRL)
         ||
            INPUT (?I56)
         end par;
         if CTRL == L then
            OUTPUT1 (I56)
         else -- assert (CTRL == N)
            par
               OUTPUT1 (I56)
            ||
               OUTPUT2 (I56)
            end par
         end if
      end loop
   end var
end process

---------------------------------------------------------------------------
-- CHOOSE_K closes the loop in the key path, by redirecting the input on
-- INPUT to OUTPUT, but for the first iteration where the original key is
-- read on FIRST_IN

process CHOOSE_K [CTRL: CP, FIRST_IN, INPUT, OUTPUT: C56] is
   var CTRL: PHASE, I56:BIT56 in
      loop
         CTRL (?CTRL);
         if CTRL == F then
            FIRST_IN (?I56)
         else -- assert (CTRL == N)
            INPUT (?I56)
         end if;
         OUTPUT (I56)
      end loop
   end var
end process

---------------------------------------------------------------------------
-- PC2 applies PC2 to generate the current subkey

process PC2 [KK: C56, SUBKEY: C48] is
   var I56: BIT56 in
      loop
	 KK (?I56);
	 SUBKEY (PC2 (I56))
      end loop
   end var
end process

end module

\end{lstlisting}

\subsection{Module DATA\_PATH}

\begin{lstlisting}[firstline=5]
module DATA_PATH (PERMUTATION_FUNCTIONS, CIPHER) is

-- DATA_PATH performs the 16 iterations ciphering DATA with the subkeys
-- read on gate SUBKEY

process DATA_PATH [DATA, OUTPUT: C64, SUBKEY: C48, CTRL_CL, CTRL_CR: CP] is
   hide FIRST_L, FIRST_R, OUTPUT_L, OUTPUT_R: C32 in
      par
         FIRST_L, FIRST_R ->
            IP [DATA, FIRST_L, FIRST_R]
      ||
         FIRST_L, FIRST_R, OUTPUT_L, OUTPUT_R ->
            hide CL_XR, CR_FX, FX_XR, XR_CR, CR_CL: C32 in
               par
                  CR_CL, CL_XR ->
                     CHOOSE_L [CTRL_CL, FIRST_L, CR_CL, CL_XR, OUTPUT_L]
               ||
                  XR_CR, CR_CL, CR_FX ->
                     CHOOSE_R [CTRL_CR, FIRST_R, XR_CR, CR_CL, CR_FX,
                               OUTPUT_R]
               ||
                  CR_FX, FX_XR ->
                     CIPHER [SUBKEY, CR_FX, FX_XR]
               ||
                  CL_XR, FX_XR, XR_CR ->
                     XOR_32 [CL_XR, FX_XR, XR_CR]
               end par
            end hide
      ||
         OUTPUT_L, OUTPUT_R ->
            IIP [OUTPUT_L, OUTPUT_R, OUTPUT]
      end par
   end hide
end process

---------------------------------------------------------------------------
-- IP applies the initial permutation IP to the initial 64-bit vector
-- received on gate DATA and breaks the resulting 64-bit vector into
-- two 32-bit vectors L and R

process IP [DATA: C64, FIRST_L, FIRST_R: C32] is
   var I64: BIT64 in
      loop
	 DATA (?I64);
	 I64 := IP (I64);
	 par
	    FIRST_L (1TO32 (I64))
	    ||
	    FIRST_R (33TO64 (I64))
	 end par
      end loop
   end var
end process

---------------------------------------------------------------------------
-- CHOOSE_L reads a 32-bit vector from INPUT and outputs to OUTPUT, but for
-- the first iteration, where it reads from FIRST_IN, and the last
-- iteration, where it outputs to LAST_OUT

process CHOOSE_L [CTRL: CP, FIRST_IN, INPUT, OUTPUT, LAST_OUT: C32] is
   var CTRL: PHASE, L32: BIT32 in
      loop
         CTRL (?CTRL);
         case CTRL in
           F ->
               FIRST_IN (?L32);
               OUTPUT (L32)
         | N ->
               INPUT (?L32);
               OUTPUT (L32)
         | L ->
               INPUT (?L32);
               LAST_OUT (L32)
         end case
      end loop
   end var
end process

---------------------------------------------------------------------------
-- CHOOSE_R reads a 32-bit vector from INPUT and outputs to OUT1 and OUT2,
-- but for the first iteration, where it reads from FIRST_IN, and the last
-- one, where it outputs to LAST_OUT

process CHOOSE_R [CTRL: CP, FIRST_IN, INPUT, OUT1, OUT2, LAST_OUT: C32] is
   var CTRL: PHASE, R32: BIT32 in
      loop
         CTRL (?CTRL);
         case CTRL in
           F ->
               FIRST_IN (?R32);
               par
                  OUT1 (R32)
               ||
                  OUT2 (R32)
               end par
         | N ->
               INPUT (?R32);
               par
                  OUT1 (R32)
               ||
                  OUT2 (R32)
               end par
         | L ->
               INPUT (?R32);
               LAST_OUT (R32)
         end case
      end loop
   end var
end process

---------------------------------------------------------------------------
-- XOR_32 asynchronously reads two 32-bit vectors and outputs their bitwise
-- sum

process XOR_32 [A, B, R: C32] is
   var A32, B32: BIT32 in
      loop
         par
            A (?A32)
         ||
            B (?B32)
         end par;
         R (XOR (A32, B32))
      end loop
   end var
end process

---------------------------------------------------------------------------
-- IIP assembles the two 32-bit vectors computed by the algorithm to a
-- 64-bit vector and applies the inverse initial permutation IIP to compute
-- the final result

process IIP [OUTPUT_L, OUTPUT_R: C32, OUTPUT: C64] is
   var OL, OH: BIT32 in
      loop
	 par
	    OUTPUT_L (?OL)
	 ||
	    OUTPUT_R (?OH)
	 end par;
	 OUTPUT (IIP (MK_64 (OH, OL)))
      end loop
   end var
end process

end module

\end{lstlisting}

\subsection{Module CIPHER}

\begin{lstlisting}[firstline=5]
module CIPHER (CHANNELS, S_BOX_FUNCTIONS) is

-- processes implementing the cipher function according to Fig. 2 of [DES]

---------------------------------------------------------------------------
-- CIPHER computes "F (R, K) = P (Si (E (R) + K))"

process CIPHER [K: C48, R, PX: C32] is
   hide ER: C48,
        IS1, IS2, IS3, IS4, IS5, IS6, IS7, IS8: C6,
        SO1, SO2, SO3, SO4, SO5, SO6, SO7, SO8: C4
   in
      par
         ER ->
            E [R, ER]
      ||
         ER, IS1, IS2, IS3, IS4, IS5, IS6, IS7, IS8 ->
            XOR_48 [ER, K, IS1, IS2, IS3, IS4, IS5, IS6, IS7, IS8]
      ||
         IS1, IS2, IS3, IS4, IS5, IS6, IS7, IS8,
         SO1, SO2, SO3, SO4, SO5, SO6, SO7, SO8 ->
            par
               S1 [IS1, SO1]
            ||
               S2 [IS2, SO2]
            ||
               S3 [IS3, SO3]
            ||
               S4 [IS4, SO4]
            ||
               S5 [IS5, SO5]
            ||
               S6 [IS6, SO6]
            ||
               S7 [IS7, SO7]
            ||
               S8 [IS8, SO8]
            end par
      ||
         SO1, SO2, SO3, SO4, SO5, SO6, SO7, SO8 ->
            P [SO1, SO2, SO3, SO4, SO5, SO6, SO7, SO8, PX]
      end par
   end hide
end process

---------------------------------------------------------------------------
-- E expands a 32-bit word to a 48-bit word using function E

process E [INPUT: C32, OUTPUT: C48] is
   var I32: BIT32 in
      loop
         INPUT (?I32);
         OUTPUT (E(I32))
      end loop
   end var
end process

---------------------------------------------------------------------------
-- XOR_48 asynchronously reads two 48-bit vectors and output their bitwise
-- sum, splitted into eight 6-bit vectors

process XOR_48 [A, B: C48, R1, R2, R3, R4, R5, R6, R7, R8: C6] is
   var A48, B48, I48: BIT48 in
      loop
         par
            A (?A48)
         ||
            B (?B48)
         end par;
         I48 := XOR (A48, B48);
         par
            R1 (1TO6   (I48))
           ||
            R2 (7TO12  (I48))
           ||
            R3 (13TO18 (I48))
           ||
            R4 (19TO24 (I48))
           ||
            R5 (25TO30 (I48))
           ||
            R6 (31TO36 (I48))
           ||
            R7 (37TO42 (I48))
           ||
            R8 (43TO48 (I48))
         end par
      end loop
   end var
end process

---------------------------------------------------------------------------

process S1 [INPUT: C6, OUTPUT: C4] is
   var I6: BIT6 in
      loop
         INPUT (?I6);
         OUTPUT (NAT_TO_BIT4 (S1[GET_ROW (I6)][GET_COLUMN (I6)]))
      end loop
   end var
end process

---------------------------------------------------------------------------

process S2 [INPUT: C6, OUTPUT: C4] is
   var I6: BIT6 in
      loop
         INPUT (?I6);
         OUTPUT (NAT_TO_BIT4 (S2[GET_ROW (I6)][GET_COLUMN (I6)]))
      end loop
   end var
end process

---------------------------------------------------------------------------

process S3 [INPUT: C6, OUTPUT: C4] is
   var I6: BIT6 in
      loop
         INPUT (?I6);
         OUTPUT (NAT_TO_BIT4 (S3[GET_ROW (I6)][GET_COLUMN (I6)]))
      end loop
   end var
end process

---------------------------------------------------------------------------

process S4 [INPUT: C6, OUTPUT: C4] is
   var I6: BIT6 in
      loop
         INPUT (?I6);
         OUTPUT (NAT_TO_BIT4 (S4[GET_ROW (I6)][GET_COLUMN (I6)]))
      end loop
   end var
end process

---------------------------------------------------------------------------

process S5 [INPUT: C6, OUTPUT: C4] is
   var I6: BIT6 in
      loop
         INPUT (?I6);
         OUTPUT (NAT_TO_BIT4 (S5[GET_ROW (I6)][GET_COLUMN (I6)]))
      end loop
   end var
end process

---------------------------------------------------------------------------

process S6 [INPUT: C6, OUTPUT: C4] is
   var I6: BIT6 in
      loop
         INPUT (?I6);
         OUTPUT (NAT_TO_BIT4 (S6[GET_ROW (I6)][GET_COLUMN (I6)]))
      end loop
   end var
end process

---------------------------------------------------------------------------

process S7 [INPUT: C6, OUTPUT: C4] is
   var I6: BIT6 in
      loop
         INPUT (?I6);
         OUTPUT (NAT_TO_BIT4 (S7[GET_ROW (I6)][GET_COLUMN (I6)]))
      end loop
   end var
end process

---------------------------------------------------------------------------

process S8 [INPUT: C6, OUTPUT: C4] is
   var I6: BIT6 in
      loop
         INPUT (?I6);
         OUTPUT (NAT_TO_BIT4 (S8[GET_ROW (I6)][GET_COLUMN (I6)]))
      end loop
   end var
end process

---------------------------------------------------------------------------
-- P collects the results of the eight processes S_BOX_i (on INi) and
-- outputs them in a single transition exit; the permutation P is applied
-- in a second step when outputting the result on OUTPUT.

process P [IN1, IN2, IN3, IN4, IN5, IN6, IN7, IN8: C4, OUTPUT: C32] is
   var I1, I2, I3, I4, I5, I6, I7, I8: BIT4 in
      loop
         par
            IN1 (?I1)
         ||
            IN2 (?I2)
         ||
            IN3 (?I3)
         ||
            IN4 (?I4)
         ||
            IN5 (?I5)
         ||
            IN6 (?I6)
         ||
            IN7 (?I7)
         ||
            IN8 (?I8)
         end par;
         OUTPUT (P (MK_32 (I1, I2, I3, I4, I5, I6, I7, I8)))
      end loop
   end var
end process

end module

\end{lstlisting}

\subsection{Module DES}

The module \lstinline+DES.lnt+ defines the architecture of the asynchronous DES together with the principal process \lstinline+MAIN+ instantiating process \lstinline+DES+.

\begin{lstlisting}[firstline=5]
module DES (CONTROLLER, DATA_PATH, KEY_PATH) is

process DES [CRYPT: CB, KEY, DATA, OUTPUT: C64] is
   hide SUBKEY: C48,
        CTRL_CL, CTRL_CR: CP,
        CTRL_SHIFT: CS, CTRL_DK, CTRL_CK: CP
   in
      par 
         SUBKEY, CTRL_CL, CTRL_CR ->
            DATA_PATH [DATA, OUTPUT, SUBKEY, CTRL_CL, CTRL_CR]
      ||
         SUBKEY, CTRL_CK, CTRL_SHIFT, CTRL_DK ->
            KEY_PATH [KEY, SUBKEY, CTRL_SHIFT, CTRL_DK, CTRL_CK]
      ||
         CTRL_CL, CTRL_CR, CTRL_CK, CTRL_SHIFT, CTRL_DK ->
            CONTROLLER [CRYPT, CTRL_CL, CTRL_CR, CTRL_SHIFT, CTRL_DK,
            CTRL_CK]
      end par
   end hide
end process

---------------------------------------------------------------------------

process MAIN [CRYPT: CB, KEY, DATA, OUTPUT: C64] is
   DES [CRYPT, KEY, DATA, OUTPUT]
end process

end module

\end{lstlisting}

\subsection{Model with Concrete Bits and Environment: DES\_SAMPLE}

This model instantiates the DES in a sequential environment providing input data and checking the output.
Hence, this model can be used to directly (i.e., non compositionally) generate an LTS for the concrete (and thus also the abstract) model.
After hiding all offers and minimization for branching bisimulation, the generated LTS is the one shown in Figure~\ref{fig:abstract_LTS}.

\begin{lstlisting}[firstline=5]
module DES_SAMPLE (DES) is

-- DES_SAMPLE uses concrete bits

process MAIN_SAMPLE [CRYPT: CB, KEY, DATA, OUTPUT: C64] is
   par CRYPT, KEY, DATA, OUTPUT in
      DES [CRYPT, KEY, DATA, OUTPUT]
   ||
      ENVIRONMENT [CRYPT, KEY, DATA, OUTPUT]
   end par
end process

---------------------------------------------------------------------------
-- 64-bit constant frequently used as example for a key

function c_13345779_9BBCDFF1 : BIT64 is
   return MK_64 (0, 0, 0, 1,  0, 0, 1, 1,  0, 0, 1, 1,  0, 1, 0, 0,
                 0, 1, 0, 1,  0, 1, 1, 1,  0, 1, 1, 1,  1, 0, 0, 1,
                 1, 0, 0, 1,  1, 0, 1, 1,  1, 0, 1, 1,  1, 1, 0, 0,
                 1, 1, 0, 1,  1, 1, 1, 1,  1, 1, 1, 1,  0, 0, 0, 1)
end function

---------------------------------------------------------------------------
-- 64-bit constant frequently used as example data

function C_01234567_89ABCDEF : BIT64 is
   return MK_64 (0, 0, 0, 0,  0, 0, 0, 1,  0, 0, 1, 0,  0, 0, 1, 1,
                 0, 1, 0, 0,  0, 1, 0, 1,  0, 1, 1, 0,  0, 1, 1, 1,
                 1, 0, 0, 0,  1, 0, 0, 1,  1, 0, 1, 0,  1, 0, 1, 1,
                 1, 1, 0, 0,  1, 1, 0, 1,  1, 1, 1, 0,  1, 1, 1, 1)
end function

---------------------------------------------------------------------------
-- result of ciphering C_01234567_89ABCDEF with C_13345779_9BBCDFF1

function C_85E81354_0F0AB405 : BIT64 is
   return MK_64 (1, 0, 0, 0,  0, 1, 0, 1,  1, 1, 1, 0,  1, 0, 0, 0,
                 0, 0, 0, 1,  0, 0, 1, 1,  0, 1, 0, 1,  0, 1, 0, 0,
                 0, 0, 0, 0,  1, 1, 1, 1,  0, 0, 0, 0,  1, 0, 1, 0,
                 1, 0, 1, 1,  0, 1, 0, 0,  0, 0, 0, 0,  0, 1, 0, 1)
end function

---------------------------------------------------------------------------
-- process simulating the environment in order to close the system,
-- executing a single encryption of c_01234567_89abcdef with
-- c_13345779_9BBCDFF1

process ENVIRONMENT [CRYPT: CB, KEY, DATA, OUTPUT: C64] is
   CRYPT (true);
   KEY (c_13345779_9BBCDFF1);
   DATA (c_01234567_89abcdef);
   -- cipher of c_01234567_89abcdef with c_13345779_9BBCDFF1
   OUTPUT (c_85e81354_0f0ab405);
   stop
end process

end module

\end{lstlisting}

\subsection{Module PROPERTY\_4}
\label{sec:lnt:property_4}

\begin{lstlisting}[firstline=5]
module property_4 is

---------------------------------------------------------------------------
-- synchronization channel without any offer

channel none is
   ()
end channel

---------------------------------------------------------------------------
-- process describing a loop starting with a synchronization on gate CRYPT,
-- followed by 16 synchronization on gate SUBKEY, where the synchronization
-- on CRYPT corresponding to the next iteration can already appear after
-- 14 synchronizations on SUBKEY.
-- this process is used to verify the presence of 16 iterations in the DES.

process MAIN [CRYPT, SUBKEY: none] is
   CRYPT;
   loop
      var IT: NAT in
         for IT := 0 while IT < 14 by IT := IT + 1 loop
            SUBKEY
         end loop
      end var;
      par
         SUBKEY; SUBKEY
      ||
         CRYPT
      end par
   end loop
end process

end module

\end{lstlisting}

%% file: lotos.tex
\section{LOTOS Specification of an Asynchronous DES}
\label{sec:lotos}

\lstset{language=LOTOS}

This appendix gives the complete LOTOS models of the DES, as required by the SVL verification scenario given in Appendix~\ref{sec:svl}.
For an introduction to LOTOS, its syntax and semantics, see the international standard~\cite{ISO-8807} or one of the tutorials listed on the CADP website\footnote{\url{http://cadp.inria.fr/tutorial/\#lotos}}.
In the sequel, a LOTOS model is called a ``specification'', in conformance with the terminology of the LOTOS standard.

\subsection{Library BIT\_CONCRETE}

The definition of type \lstinline+BIT+ by library \lstinline+BIT_CONCRETE+is not strictly necessary, because one could use the library \lstinline+BIT+ provided with the CADP toolbox, which defines many more operations than just \lstinline+xor+.
Thus, the library \lstinline+BIT_CONCRETE+ is included here for better comparison with its abstract version (see Appendix~\ref{sec:lotos:abstract_bit} below).

\begin{lstlisting}[linerange=14-19]

type BIT is BOOLEAN
   sorts BIT
   opns 0 (*! implementedby BIT_ZERO constructor *) : -> BIT 
        1 (*! implementedby BIT_ONE  constructor *) : -> BIT
endtype
\end{lstlisting}

\subsection{Library BIT\_ABSTRACT}
\label{sec:lotos:abstract_bit}

This library is a replacement of the type \lstinline+BIT+, transforming the two-valued type into a singleton, which transitively transforms all bit vectors into singletons as well, thus completely abstracting data.

\begin{lstlisting}[linerange=21-28]
type BIT is BOOLEAN
   sorts BIT
   opns 0 (*! constructor *) : -> BIT 
        1 : -> BIT
   eqns
      ofsort BIT
         1 = 0;
endtype
\end{lstlisting}

\subsection{Library TYPES}

This library defines several data types modeling different sizes of bit vectors and types required for the control of the DES.

\lstinputlisting[linerange=15-469]{TYPES.lib}

\subsection{Library PERMUTATION\_FUNCTIONS}

This library defines the different permutation functions used by the DES.

\begin{lstlisting}[linerange=16-126, aboveskip=0pt]

type PERMUTATION_FUNCTIONS is BIT32, BIT48, BIT56, BIT64
   opns E   : BIT32 -> BIT48    (* expansion *)
        IP  : BIT64 -> BIT64    (* initial permutation *)
        IIP : BIT64 -> BIT64    (* inverse initial permutation *)
        P   : BIT32 -> BIT32    (* permutation *)
        PC1 : BIT64 -> BIT56    (* permuted choice 1 *)
        PC2 : BIT56 -> BIT48    (* permuted choice 2 *)
   eqns
      forall B1,  B2,  B3,  B4,  B5,  B6,  B7,  B8,
             B9,  B10, B11, B12, B13, B14, B15, B16,
             B17, B18, B19, B20, B21, B22, B23, B24,
             B25, B26, B27, B28, B29, B30, B31, B32,
             B33, B34, B35, B36, B37, B38, B39, B40,
             B41, B42, B43, B44, B45, B46, B47, B48,
             B49, B50, B51, B52, B53, B54, B55, B56,
             B57, B58, B59, B60, B61, B62, B63, B64 : BIT
      ofsort BIT48
           E (MK_32 (B1,  B2,  B3,  B4,  B5,  B6,  B7,  B8,
                     B9,  B10, B11, B12, B13, B14, B15, B16,
                     B17, B18, B19, B20, B21, B22, B23, B24,
                     B25, B26, B27, B28, B29, B30, B31, B32)) =
              MK_48 (B32, B1,  B2,  B3,  B4,  B5,
                     B4,  B5,  B6,  B7,  B8,  B9,
                     B8,  B9,  B10, B11, B12, B13,
                     B12, B13, B14, B15, B16, B17,
                     B16, B17, B18, B19, B20, B21,
                     B20, B21, B22, B23, B24, B25,
                     B24, B25, B26, B27, B28, B29,
                     B28, B29, B30, B31, B32, B1);
      ofsort BIT32
           P (MK_32 (B1,  B2,  B3,  B4,  B5,  B6,  B7,  B8,
                     B9,  B10, B11, B12, B13, B14, B15, B16,
                     B17, B18, B19, B20, B21, B22, B23, B24,
                     B25, B26, B27, B28, B29, B30, B31, B32)) =
              MK_32 (B16, B7,  B20, B21,
                     B29, B12, B28, B17,
                     B1,  B15, B23, B26,
                     B5,  B18, B31, B10,
                     B2,  B8,  B24, B14,
                     B32, B27, B3,  B9,
                     B19, B13, B30, B6,
                     B22, B11, B4,  B25);
      ofsort BIT56
         PC1 (MK_64 (B1,  B2,  B3,  B4,  B5,  B6,  B7,  B8,
                     B9,  B10, B11, B12, B13, B14, B15, B16,
                     B17, B18, B19, B20, B21, B22, B23, B24,
                     B25, B26, B27, B28, B29, B30, B31, B32,
                     B33, B34, B35, B36, B37, B38, B39, B40,
                     B41, B42, B43, B44, B45, B46, B47, B48,
                     B49, B50, B51, B52, B53, B54, B55, B56,
                     B57, B58, B59, B60, B61, B62, B63, B64)) =
              MK_56 (B57, B49, B41, B33, B25, B17, B9,
                     B1,  B58, B50, B42, B34, B26, B18,
                     B10, B2,  B59, B51, B43, B35, B27,
                     B19, B11, B3,  B60, B52, B44, B36,
                     B63, B55, B47, B39, B31, B23, B15,
                     B7,  B62, B54, B46, B38, B30, B22,
                     B14, B6,  B61, B53, B45, B37, B29,
                     B21, B13, B5,  B28, B20, B12, B4);
      ofsort BIT48
         PC2 (MK_56 (B1,  B2,  B3,  B4,  B5,  B6,  B7,  B8,
                     B9,  B10, B11, B12, B13, B14, B15, B16,
                     B17, B18, B19, B20, B21, B22, B23, B24,
                     B25, B26, B27, B28, B29, B30, B31, B32,
                     B33, B34, B35, B36, B37, B38, B39, B40,
                     B41, B42, B43, B44, B45, B46, B47, B48,
                     B49, B50, B51, B52, B53, B54, B55, B56)) =
              MK_48 (B14, B17, B11, B24, B1,  B5,
                     B3,  B28, B15, B6,  B21, B10,
                     B23, B19, B12, B4,  B26, B8,
                     B16, B7,  B27, B20, B13, B2,
                     B41, B52, B31, B37, B47, B55,
                     B30, B40, B51, B45, B33, B48,
                     B44, B49, B39, B56, B34, B53,
                     B46, B42, B50, B36, B29, B32);
      ofsort BIT64
          IP (MK_64 (B1,  B2,  B3,  B4,  B5,  B6,  B7,  B8,
                     B9,  B10, B11, B12, B13, B14, B15, B16,
                     B17, B18, B19, B20, B21, B22, B23, B24,
                     B25, B26, B27, B28, B29, B30, B31, B32,
                     B33, B34, B35, B36, B37, B38, B39, B40,
                     B41, B42, B43, B44, B45, B46, B47, B48,
                     B49, B50, B51, B52, B53, B54, B55, B56,
                     B57, B58, B59, B60, B61, B62, B63, B64)) =
              MK_64 (B58, B50, B42, B34, B26, B18, B10, B2,
                     B60, B52, B44, B36, B28, B20, B12, B4,
                     B62, B54, B46, B38, B30, B22, B14, B6,
                     B64, B56, B48, B40, B32, B24, B16, B8,
                     B57, B49, B41, B33, B25, B17, B9,  B1,
                     B59, B51, B43, B35, B27, B19, B11, B3,
                     B61, B53, B45, B37, B29, B21, B13, B5,
                     B63, B55, B47, B39, B31, B23, B15, B7);
      ofsort BIT64
         IIP (MK_64 (B1,  B2,  B3,  B4,  B5,  B6,  B7,  B8,
                     B9,  B10, B11, B12, B13, B14, B15, B16,
                     B17, B18, B19, B20, B21, B22, B23, B24,
                     B25, B26, B27, B28, B29, B30, B31, B32,
                     B33, B34, B35, B36, B37, B38, B39, B40,
                     B41, B42, B43, B44, B45, B46, B47, B48,
                     B49, B50, B51, B52, B53, B54, B55, B56,
                     B57, B58, B59, B60, B61, B62, B63, B64)) =
              MK_64 (B40, B8,  B48, B16, B56, B24, B64, B32,
                     B39, B7,  B47, B15, B55, B23, B63, B31,
                     B38, B6,  B46, B14, B54, B22, B62, B30,
                     B37, B5,  B45, B13, B53, B21, B61, B29,
                     B36, B4,  B44, B12, B52, B20, B60, B28,
                     B35, B3,  B43, B11, B51, B19, B59, B27,
                     B34, B2,  B42, B10, B50, B18, B58, B26,
                     B33, B1,  B41, B9,  B49, B17, B57, B25);
endtype
\end{lstlisting}

\subsection{Library S\_BOX\_FUNCTIONS}

Contrary to the LNT model, the LOTOS model defines each S-box as a functions, which completely expresses the \emph{function} of the corresponding S-box, rather than the array defining the S-box.
This avoids the definition of additional data types for the array, together with the accessor functions (which are generated automatically by the LNT2LOTOS translator).
Precisely, each of the following S-box functions associates to any 6-bit vector the 4-bit value encoding of the corresponding entry in the associated S-box table (see Appendix~\ref{sec:lnt:s_boxes} for an encoding of these tables).
The DES standard defines that for a 6-bit vector $b_1b_2b_3b_4b_5b_6$, corresponds the entry in the $b_1b_6$-th row and the $b_2b_3b_4b_5$-th column.
The defining equations are ordered such that the right hand sides correspond to the usual reading order of the S-box tables of the standard (rows from top to bottom, and each row from left to right).

The definitions of these functions use premisses, because in the case type \lstinline+BIT+ implements the abstract singleton type, \lstinline+1+ is no constructor and thus not supported on the left hand side of an equation, e.g., the equation
\begin{lstlisting}
         S1 (MK_6 (0, 0, 0, 0, 1, 0) = MK_4 (0, 1, 0, 0);
\end{lstlisting}
would be rejected by the LOTOS compiler.

\lstinputlisting[linerange=16-553, aboveskip=0pt]{S_BOX_FUNCTIONS.lib}

\subsection{Library CONTROLLER}

Process \lstinline+CONTROLLER+ is responsible for the generation of the appropriate commands to the shift register of the key path as well as the four multiplexers of the data and key paths.
It consists of a parallel composition of six processes: one process per controlled block (multiplexer or shift register) plus a process counting the iterations.

\lstinputlisting[linerange=23-107]{CONTROLLER.lib}

\subsection{Library KEY\_PATH}

Process \lstinline+KEY_PATH+ describes the architecture of the key path, generating the sixteen subkeys required for the sixteen iterations.

\lstinputlisting[linerange=16-156]{KEY_PATH.lib}

\subsection{Library CIPHER}

The processes of library \lstinline+CIPHER.lib+ implement Figure~2 of the standard~\cite{FIPS-46-3}, i.e., they compute \[P \Bigl(S_i \bigl(E (R_i) + K_i\bigr)\Bigr)\] where $K_i$ is the $i$-th subkey and $R_i$ is the 32-bit vector handled by the $i$-th iteration of the DES.

\lstinputlisting[linerange=19-214]{CIPHER.lib}

\subsection{Library DATA\_PATH}

This library defines the process \lstinline+DATA_PATH+, performing the 16 iterations of the DES, outputing on gate \lstinline+OUTPUT+ the result of (de)ciphering the data read on gate \lstinline+DATA+ using the sequence of 16 subkeys received on gate \lstinline+SUBKEY+.

\lstinputlisting[linerange=17-160]{DATA_PATH.lib}

\subsection{Library DES}

The library \lstinline+DES.lib+ factorizes the definition of the architecture of the asynchronous DES, and is shared by all three LOTOS specifications of the DES, namely \lstinline+DES_ABSTRACT.lotos+, \lstinline+DES_CONCRETE.lotos+, and \lstinline+DES_SAMPLE.lotos+.

\begin{lstlisting}[linerange=16-26]

process DES [CRYPT, KEY, DATA, OUTPUT] : noexit :=
   hide SUBKEY, CTRL_CL, CTRL_CR, CTRL_CK, CTRL_SHIFT, CTRL_DK in
      (
         DATA_PATH [DATA, OUTPUT, SUBKEY, CTRL_CL, CTRL_CR]
         |[SUBKEY]|
         KEY_PATH [KEY, SUBKEY, CTRL_CK, CTRL_SHIFT, CTRL_DK]
      )
      |[CTRL_CL, CTRL_CR, CTRL_CK, CTRL_SHIFT, CTRL_DK]|
      CONTROLLER [CRYPT, CTRL_CL, CTRL_CR, CTRL_CK, CTRL_SHIFT, CTRL_DK]
endproc
\end{lstlisting}

\subsection{Specification with Abstract Bits: DES\_ABSTRACT}

This LOTOS specification can be used for compositional state space generation and verification.
Due to the abstraction, there is no need to close the system with an environment.

\lstinputlisting[linerange=15-33]{DES_ABSTRACT.lotos}

\subsection{Specification with Concrete Bits: DES\_CONCRETE}

This LOTOS specification can be used to generate a prototype implementation.
It is not suitable for state space generation, because it would have to enumerate over all possible input values, i.e., 64-bit vectors.

\lstinputlisting[linerange=15-33]{DES_CONCRETE.lotos}

\subsection{Specification with Concrete Bits and Environment: DES\_SAMPLE}

This LOTOS specification can be used to directly (i.e., non compositionally) generate an LTS.
After hiding all offers and minimization for branching bisimulation, the generated LTS is the one shown in Figure~\ref{fig:abstract_LTS}.

\lstinputlisting[linerange=15-79]{DES_SAMPLE.lotos}

\subsection{Specification of Property~4}
\label{sec:lotos:property_4}

This LOTOS specification describes an automaton, which is used to verify the correct synchronisation of the data and key paths by equivalence checking.

\lstinputlisting[linerange=14-68]{property_4.lotos}

%% file: c.tex
\section{C Code for the EXEC/C\AE{}SAR Framework}
\label{sec:execcaesar_c}

\lstset{language=C}

To generate a protoype from the LNT or LOTOS model using the EXEC/C\AE{}SAR framework, additional C code is necessary, namely the main program (for which the example provided with CADP\footnote{Precisely, the file \texttt{\$CADP/src/exec\_caesar/main.c} included in the CADP toolbox.} can be used) and so-called \emph{gate functions} implementing the interaction with the environment.
The DES prototype reads its inputs from the standard input and prints its results to the standard output.
Each rendezvous corresponds to one line of input (respectively, output), following the syntax of LOTOS, i.e., ``\lstinline[language=LOTOS]+G !O+'', where \lstinline[language=LOTOS]+G+ is a gate name and \lstinline[language=LOTOS]+O+ is a sequence of characters corresponding to the offer.
For convenience, 64-bit vectors (for gates  \lstinline[language=LNT]+CRYPT+, \lstinline[language=LNT]+DATA+, \lstinline[language=LNT]+KEY+, and \lstinline[language=LNT]+OUTPUT+) are represented by a sequence of sixteen hexadecimal digits, and Booleans (for gate \lstinline[language=LOTOS]+CRYPT+) by \lstinline[language=LOTOS]+0+ (false) and  \lstinline[language=LOTOS]+1+ (true).
The prototype can also write its execution trace in the SEQ format\footnote{\url{http://cadp.inria.fr/man/exhibitor.html\#sect3}} to a log file.

\subsection{Auxiliary Functions for Reading and Writing}

The following auxiliary functions are required to parse and print 64-bit vectors.
Each line of input parsed is stored in a variable (local to the module defining the gate functions), because the first gate function called by the prototype might not correspond to the input line, enabling its reuse by another gate function.

\lstinputlisting[linerange=40-228]{gate_functions.c}

\subsection{Gate Functions}

\definecolor{grey}{RGB}{160,160,160}
\lstset{moredelim=[il][\footnotesize\color{grey}]{@}}

Lines typeset in {\footnotesize\color{grey}small, light grey font} correspond to automatically generated code to check assumptions about the parameters of gate functions and (possibly) log gate function calls to a file.
These functions are documented in the header file \lstinline+caesar_kernel.h+ included in the CADP toolbox.

\lstinputlisting[firstline=230]{gate_functions.c}
\lstset{deletedelim=[il]{@}}

%% file: svl.tex
\section{Complete Verification Scenarios for the Asynchronous DES}
\label{sec:svl}

\lstset{language=SVL}

The complete verification scenario of the asynchronous DES is executed by the following SVL script, using only sequential tools of CADP (i.e., no distributed state space generation is required\footnote{The screenshots used in the manual page of the DISTRIBUTOR tool where obtained by the dsitributed generation of \lstinline[language=LOTOS]+DES_SAMPLE.lotos+}).
The script is also available in the demo example on the CADP website\footnotemark[\value{demo38}].

The SVL script first compositionally generates the LTS corresponding to the abstract LNT model \lstinline[language=LNT]+DES_ABSTRACT.lnt+.
The script then verifies several properties of the LNT models, using different techniques, such as model checking (for the temporal logic properties of Sections~\ref{sec:abstract} and \ref{sec:concrete}) and equivalence checking, but also the generation of a prototype from the LNT model \lstinline[language=LNT]+DES_CONCRETE.lnt+ and comparing its output for some example data and key with official results.
These first steps can be easily adapted to use the LOTOS model instead of the LNT model.
Finally, the LNT models are compared (using equivalence checking) with the corresponding LOTOS specifications, which are supposed to be located in a subdirectory called ``\texttt{LOTOS}'', together with an SVL script called ``\texttt{demo.svl}'' generating the LTSs for the LOTOS specifications \lstinline[language=LOTOS]+DES_ABSTRACT.lotos+ and \lstinline[language=LOTOS]+DES_SAMPLE.lotos+.

For a description of the syntax of SVL, see the SVL manual page\footnote{\url{http://cadp.inria.fr/man/svl.html}} --- the most important points being that lines starting with ``\lstinline+%+'' are Bourne shell commands, and that temporal logic formulas in MCL (Model Checking Language)~\cite{Mateescu-Thivolle-08} are directly inlined.

\lstinputlisting[firstline=18]{demo.svl}